# Quantitative assessment of perturbation theory-based lattice thermal conductivity models using quasi-continuum approximation

Ahmed Hamed[1,2], Anter El-Azab[2,1]

[1]School of Nuclear Engineering, Purdue University, West Lafayette, IN 47907, USA

[2]School of Materials Engineering, Purdue University, West Lafayette, IN 47907, USA

**Abstract**

The impact of dispersion relations, anisotropy, and Brillouin zone structure on intrinsic phonon scattering rates has been assessed within the harmonic approximation–perturbation theory approach for lattice dynamics. An anisotropic nonlinear elastic continuum has been considered with various levels of representation of phonon dispersion and Brillouin zone shape, and with Grüneisen parameter used as an average measure of crystal anharmonicity. In addition, thermal conductivity prediction of different models for the treatment of the off-diagonal elements of phonon collision operator are compared. For a model system, argon, with a relatively high anisotropy ratio, the results show that accounting for anisotropy is critical for accurate determination of the available phase space for 3-phonon scattering and the scattering rates. In addition, widely spread approximations such as isotropic continuum and Single Mode Relaxation Time (SMRT) are found unreliable, even for the case of cubic systems. The success of these approximations is demonstrated to be a direct result of error cancellations. By benchmarking against our iterative solution of Boltzmann Transport Equation, which achieves excellent agreement with experimental thermal conductivity data for solid argon (2–80 K), we show the essential importance of considering coupling terms of the phonon scattering kernel at phonon mode level, and not in a statistical average sense as, for example, Callaway's model does. Moreover, our results manifest the significance of the role played by coherent phonon scattering near the melting temperature, in agreement with molecular dynamics findings; which serves as an evidence for the crossover between the heat diffusion mediated by particle-like phonons (incoherent scattering) and the wave-like heat propagation due to phonon coherent scattering. Furthermore, sensitivity of conductivity prediction to phonon spectrum is revealed to change over temperature.

*Keywords:* Lattice thermal conductivity, Phonon Boltzmann transport equation, Perturbation theory.

# 1. Introduction

This communication assesses and compares common approximations for intrinsic lattice thermal conductivity computation using phonon Boltzmann Transport Equation (BTE). In particular, the effects of dispersion anisotropy and relations, Brillouin zone structure, and the correlations between the 3-phonon scattering rates of different normal modes on thermal conductivity prediction are evaluated within the perturbation theory/linearized BTE framework. This is motivated by the increasing interest to develop reliable models for theoretical prediction of materials thermal performance where phonons play a major role. Such models help to design new semiconductors and dielectric materials and components with enhanced properties and optimize the performance of thermoelectrics, nanoelectronics, and thermal therapy devices [1-9].

Phenomenological heat diffusion models are suitable for diffusion regime, where the mean free path of the heat carriers is much shorter than the characteristic physical dimension of the system. At sub-micron scale, however, phonon transport based models must be used to understand heat transfer in semiconductors and dielectric crystals. Unlike diffusion models, phonon transport models can capture the microstructure effects on thermal transport [10-13] and incorporate simultaneously different modes of energy propagation. Such models are useful in situations where the system under consideration embraces broad spectrum of phonon mean free paths with different modes of transport, i.e., involving diffusive and ballistic regimes. Transport properties, in linear response regime, can be extracted from the statistical fluctuations in equilibrium, using fluctuation-dissipation theorem. Thermal conductivity can thus be evaluated using Green-Kubo relations by calculating heat flux autocorrelation function [14-18]. This approach yields a reasonable prediction for thermal conductivity, as it can take full account of crystal anharmonicity. However, it is computationally expensive and may suffer significant uncertainties in some cases. Computational cost imposes constrains on the size of simulation cells, and hence the number of accessible microstates. In addition, classical molecular dynamic is not reliable below Debye temperature, as it does not account for quantum effects, for example, it assumes equal contributions to the specific heat for each internal degree of freedom of vibrational modes and ignores the quantization of vibrational modes energy. Bedoya-Martínez et al. [14] reported on the failure of classical molecular dynamic to reproduce experimental thermal conductivity of argon at low temperature, even when quantum thermostat was invoked, and concluded that this problem is still an open challenge.

At the same time, phonon BTE is one of the standard techniques for the theoretical prediction of lattice thermal conductivity and studying heat flow at mesoscale [19-21]. The quantum mechanical perturbation theory [22,23], within the picture of weakly interacting normal modes, provides a mechanistic framework for the evaluation of the phonon collision kernel of the scattering term in BTE. For tractability reasons, the linearized form of BTE has been widely used. However, solving the linearized BTE is an intricate problem. Several relevant computational models were developed, the main differences among which pertain to the input parameters and the correlation between the scattering rates of different phonon normal modes. The input parameters include the harmonic (dispersion relations) and anharmonic Interatomic Force Constants (IFC) and Brillouin Zone (BZ) structure. For the treatment of correlations between phonon normal modes, different methods have been used to solve this system of coupled linear equations, e.g., Single Mode Relaxation Time (SMRT) approximation [20,21,24], Callaway's model [10,25], variational methods [10,20], and iterative scheme (the most reliable among these) [20,26]. More recently, by diagonalizing the scattering operator of BTE, Cepellotti and Marzari [27] proposed "relaxons" (defined as collective phonon excitations) as a more adequate representations of heat carriers, through crystal lattice, in kinetic theory with well-defined relaxation times. They pointed out that this transformation, from phonon to relaxon coordinates, is more important in regimes where normal processes are prominent, i.e., at cryogenic temperatures and in the case of two-dimensional materials. These remarks emphasize the importance of accurate modelling of the interplay between normal and umklapp processes for reliable representation of intrinsic lattice thermal resistivity.

A steady state, spatially homogenous solution of BTE is traditionally employed to yield bulk thermal conductivity directly using kinetic theory and linear response theory [10,20]. In kinetic theory, the expression for thermal conductivity (presented later in Sec. 2.5) involves a double integral/sum over momentum space of phonon normal modes. Phonon frequency spectrum appears twice in this double integral, once in the outer and once in the inner integrals. The outer integral adds the individual contribution of phonon normal modes of different polarizations to lattice thermal conductivity. In this integral, there is no explicit dependency on the wavevector or the polarization eigenvector of individual normal modes. On the other hand, the inner integral depends explicitly on the wavevectors and the polarization eigenvectors of the interacting phonon triplets (in addition to their eigenvalues and the temperature, as in the outer integrand). It evaluates the scattering rates of each phonon normal mode, by adding up the contribution of individual scattering events to the scattering strength. This inner integral yields the effective phonon life times as well. Consequently, for any numerical investigation, to assess the influence of the used phonon states structure on the predictability of thermal conductivity, the impact of the same structure on the scattering rates should be sought mechanistically. Mixing phenomenological models for relaxation times such as Holland's model [28,29], which is still in use today, with real phonon spectrum makes the final prediction questionable. For example, isotropic continuum approximation inherently underestimates the available phase space for 3-phonon scattering; however, it introduces Van Hove singularity to the phonon spectrum and hence exaggerates the fraction of phonon modes with low or zero group velocity. This effect is marginally offset by assuming linear dispersion. Accordingly, it is impossible to qualitatively anticipate the overall effect of these different approximations on thermal conductivity. In addition, some approximations are good only at certain range of temperature (e.g., Debye model at high temperature) and their behavior out of their range of validity is unknown. Nevertheless, several numerical studies inadequately apply these approximations at other ranges. In the absence of quantitative assessment of the associated error, the predictability and reliability of these models will remain unclear, and error cancellation and/or the use of adjustable parameters will be always doubted as the reason of their success.

In this paper, we present the computational details of our robust algorithm for intrinsic lattice thermal conductivity computation. This includes direct incorporation of the dispersion curves in high symmetry directions, to account for cubic anisotropy of dispersion relations for accurate evaluation of 3-phonon phase space, which per se can be coupled to a Monte Carlo solver of BTE (instead of the widely used isotropic dispersion assumption). In addition, the model employs an adjustable-parameter-free implementation of Dirac delta function and the iterative scheme to solve BTE. By applying this model to Mie—Lennard–Jones argon [30], we captured successfully for the first time the characteristic $T^2$ behavior of argon experimental thermal conductivity at low temperature, T, and the peak at temperature of 8 K, in addition to the classical $T^{-1}$ behavior above 20 K, by the sole use of 3-phonon scattering. We use this model, which is in excellent agreement with experiment over the entire temperature range (2–80 K), to benchmark the considered approximations, which are summarized in Table 1 in Sec. 4. Our goal of the current study is twofold. *First*, to assess the validity of isotropic continuum approximation for cubic single crystal of materials with high Zenter anisotropy ratio, $A = 2c_{44}/(c_{11}-c_{12}) > 2$, and to evaluate the sensitivity of thermal conductivity prediction to dispersion relation in high index crystallographic directions. In this regard, we demonstrate that the departure from 1/T behavior of thermal conductivity at high temperature, under harmonic approximation, reported in several computational studies is partially because of isotropic continuum assumption. *Second*, to emphasize the role played by the collective relaxation of phonon modes and asses different models accounting for the off-diagonal elements of the phonon collision operator. At high temperature, due to phonon collective scattering, our results illustrate the crossover between the heat diffusion mediated by particle-like phonons and the wave-like heat propagation [31-33]. Moreover, the results exhibit the importance of considering the correlation between the relaxation of different phonon modes at mode-level, and not in a statistical average sense as Callaway's model does, particularly at low temperature. SMRT approximation is shown to be inadequate for argon, and anisotropy in thermal conductivity of cubic argon is predicted due to phonon focusing [34-35]. Finally, this study manifest quantitatively the existence of characteristic peak thermal conductivity at finite temperature in perfect crystals, regardless of the dispersion and relaxation times models used.

## 2. Theory

### 2.1. Phonon transport and Linearized BTE

Formally speaking, the transformation of the atomic vibration problem to phonon normal modes analysis is a plane wave expansion. Accordingly, any rigorous treatment of phonon modes propagation should seek their wave nature, e.g., Green's function methods [5,10]. However, for many practical purposes, phonon wavelengths are small enough that their wave effect on a discrete atomistic lattice does not need to be considered explicitly [5,20]. In such situations, where non-local effects are not important, a quasi-particle picture can be invoked, where wave effects are incorporated into the scattering term. Hence, thermal transport can be studied by tracking the temporal evolution of phonon population in phase space using the semi-classical phonon BTE, which takes the form

$$\frac{\partial n_{qs}}{\partial t} + \boldsymbol{v}_g^{qs} \cdot \nabla n_{qs} = \left.\frac{\partial n_{qs}}{\partial t}\right|_C. \tag{1}$$

The fundamental assumption in the derivation of this equation is that there exist a distribution function $n_{qs}(r,t)$ which measures the occupation number of phonons in mode $qs$, where $q$ stands for the phonon wavevector and $s$ labels polarization branch index, in the neighborhood of point $r$ (in the real space) at time $t$. In Eq. (1), $\boldsymbol{v}_g^{qs}$ denotes the group velocity of the mode $qs$. The local change of this distribution with respect to time is attributed to two mechanisms: phonon drift term, expressed by the second term to the LHS of Eq. (1), and phonon scattering (collision), expressed by the term to the RHS. The spatial gradient of phonon distribution is ascribed to the existence of temperature gradient, while phonon scattering by phonon interaction processes acts as the restoring term to equilibrium distribution.

Solving BTE exactly is a formidable task, basically because the scattering term requires knowledge of all transition rates and phonon occupation numbers of all phonon modes. A simplification can be made, by noting that at steady state, and in the presence of small temperature gradient, the deviation from equilibrium distribution is so small that we can expand phonon populations about their thermodynamic equilibrium distribution $\bar{n}_{qs}$ at temperature $T$, which follows Bose-Einstein distribution with zero chemical potential. Accordingly, for a given normal mode with frequency $\omega_{qs}$, the equilibrium distribution can be obtained from

$$\bar{n}_{qs} = 1/[\exp(\hbar\omega_{qs}/k_B T) - 1], \tag{2}$$

in which $k_B$ and $\hbar$ are Boltzmann constant and the reduced Plank constant, respectively. Retaining the linear term of Taylor-series expansion, phonon occupation number is given by

$$n_{qs} \simeq \bar{n}_{qs} + n'_{qs} = \bar{n}_{qs} - \psi_{qs}\frac{\partial \bar{n}_{qs}}{\partial(\hbar\omega_{qs})} = \bar{n}_{qs} + \frac{1}{k_B T}\psi_{qs}\bar{n}_{qs}(\bar{n}_{qs}+1), \tag{3}$$

where $\psi_{qs}$ is a scalar distribution function measures the deviation of phonon occupation number from the equilibrium distribution. When phonon modes are weakly interacting and nonlocality effects are negligible, the spatial gradient of the drift term can be expressed in terms of the local temperature derivative and the temperature gradient using chain rule. Moreover, the deviational term can be neglected. Based on that, the drift term in the linearized form of BTE is expressed as

$$\boldsymbol{v}_g^{qs} \cdot \nabla n_{qs} \simeq \boldsymbol{v}_g^{qs} \cdot \nabla T \frac{\partial \bar{n}_{qs}}{\partial T} = \frac{\hbar \omega_{qs}}{k_B T^2} \boldsymbol{v}_g^{qs} \cdot \nabla T \bar{n}_{qs} (\bar{n}_{qs} + 1). \tag{4}$$

This approximation produces what is called linearized Boltzmann equation with a canonical form: X= Pψ, where X is the inhomogeneity created by the temperature gradient, and P is phonon collision operator. The form of phonon collision operators depends on the type of the phonon scattering (elastic or inelastic) and consists of two components, namely, diagonal and off-diagonal terms.

The linearized semi-classical BTE can be used to predict thermal conductivity in two ways. The first is to solve this integro-differential equation for phonons numerically to yield the heat flux directly from phonon transport. In this case, the conductivity is estimated by selecting idealized situations involving temperature gradients in specific directions, which, together with the heat flux computed from phonon transport, are used to determine the conductivity by inverting Fourier law of heat conduction [19]. For this purpose, linearized BTE under relaxation time approximation (to be discussed in Sec. 2.3.) can be solved stochastically, using Monte Carlo technique, or deterministically using discrete ordinate, spectral, or finite volume methods [5]. Additionally, phonon kinetic theory may be used for the direct solution of space-homogenous BTE under relaxation time approximation. In this case, steady state BTE for a given normal mode takes on the form

$$\frac{\hbar \omega_{qs}}{k_B T^2} \boldsymbol{v}_g^{qs} \cdot \nabla T \bar{n}_{qs} (\bar{n}_{qs} + 1) = -\frac{n_{qs} - \bar{n}_{qs}}{\tau_{qs}}, \tag{5}$$

with $\tau_{qs}$ designates the relaxation time of the phonon mode $qs$.

## 2.2. Harmonic approximation—Perturbation theory approach

The time dependent perturbation theory provides first-order transition rates between harmonic phonon states induced by crystal anharmonicity [10-13,22]. This is done using planar wave expansion to solve Schrödinger equation in terms of the eigenfunctions of the harmonic component of the quantum mechanical crystal Hamiltonian, with a perturbation accounting for crystal anharmonicity. The theory thus aimed to evaluate the elements of transition probability matrix in state space associated with small anharmonic perturbations of the non-interacting harmonic vibrational states, which are obtained by applying second quantization scheme for coordinate transformation [10-13,23]. In this regard, it is worth to mention that the validity of the perturbative solution is contingent on small relative values of the amplitude of atomic displacement with respect to the interatomic spacing and the frequency shift and width with respect to the harmonic frequency [10,23]. Otherwise, different method should be sought, e.g., phonon self-consistent method [10,36,37].

Retaining the leading term of crystal anharmonicity, the interaction between normal modes is possible through 3-phonon processes. In this view, Fermi golden rule can be used to evaluate the probability of three-phonon scattering events, which are governed by energy conservation and momentum selection rules. The transition rates are expressed explicitly in terms of phonon frequencies and eigenvectors (harmonic IFCs), the anharmonic IFCs, and phonon ladder operators. Based on phonon ladder operators, we can have either fission (creation) or fusion (annihilation) interaction. Another classification can be made based on quasi-momentum conservation of phonon triplet wavevectors. So, processes that involve reciprocal lattice vector (G) to achieve momentum conservation are called umklapp processes, while others that conserve the momentum on their own (i.e., G = 0) are called normal processes. Accordingly, the intrinsic collision term due to 3-phonon processes is given by

$$-\left(\frac{\partial n}{\partial t}\right)_C = \sum_{q's',q''s''} [(P_{qs,q's'}^{q''s''} - P_{q''s''}^{qs,q's'}) + \frac{1}{2}(P_{qs}^{q's',q''s''} - P_{q's',q''s''}^{qs})]. \qquad (6)$$

Using Fermi golden rule, the transition rate for fusion, $P_{qs,q's'}^{q''s''}$, and fission, $P_{qs}^{q's',q''s''}$, events can be expressed, respectively, as

$$P_{qs,q's'}^{q''s''} = \frac{2\pi}{\hbar^2}\left|\langle n_{qs}-1, n_{q's'}-1, n_{q''s''}+1|H'|n_{qs}, n_{q's'}, n_{q''s''}\rangle\right|^2 \times \delta(\omega(q''s'')-\omega(qs)-\omega(q's')), \qquad (7a)$$

$$P_{qs}^{q's',q''s''} = \frac{2\pi}{\hbar^2}\left|\langle n_{qs}-1, n_{q's'}+1, n_{q''s''}+1|H'|n_{qs}, n_{q's'}, n_{q''s''}\rangle\right|^2 \times \delta(\omega(q''s'')-\omega(qs)+\omega(q's')). \qquad (7b)$$

Here, $\delta(\omega(q''s'')-\omega(qs)\pm\omega(q's'))$ is Dirac delta function, which signifies that the transition from an initial eigenstate with energy $E_i$ to a final eigenstate with energy $E_f$ through 3-phonon scattering process is governed by the rule of energy conservation in statistical average sense, where: $E_f - E_i = \hbar(\omega(q''s'')-\omega(qs)\pm\omega(q's'))$. From the properties of phonon ladder operators [10-13,23], Eqs. (7a) and (7b) can be simplified respectively to

$$P_{qs,q's'}^{q''s''} = \frac{2\pi}{\hbar^2}\left|\Phi^{qs,q's',q''s''}\right|^2 \times n_{qs}n_{q's'}(n_{q''s''}+1)\delta(\omega_{qs}+\omega_{q's'}-\omega_{q''s''})\delta_{q+q'+q'',G}, \qquad (8a)$$

$$P_{qs}^{q's',q"s"} = \frac{2\pi}{\hbar^2}\left|\Phi^{qs,q's',q"s"}\right|^2 \times (n_{qs}+1)n_{q's'}n_{q"s"}\delta(\omega_{qs}-\omega_{q's'}-\omega_{q"s"})\delta_{q-q'+q",G}. \tag{8b}$$

By applying the principle of microscopic reversibility (detailed balance) and the linearization technique discussed before, we can express the scattering term in terms of the equilibrium distributions and 3-phonon transition rates ($\overline{P}_{3ph}$), in addition to the deviation from equilibrium term [10]:

$$-\left.\frac{\partial n_{qs}}{\partial t}\right|_{3ph} = \frac{1}{k_B T}\sum_{q's',q"s"}[\overline{P}_{qs,q's'}^{q"s"}(\psi_{qs}+\psi_{q's'}-\psi_{q"s"})$$
$$+\frac{1}{2}\overline{P}_{qs}^{q's',q"s"}(\psi_{qs}-\psi_{q's'}-\psi_{q"s"})], \tag{9}$$

where

$$\overline{P}_{qs,q's'}^{q"s"} = \frac{2\pi}{\hbar^2}\left|\Phi^{qs,q's',q"s"}\right|^2 \times \overline{n}_{qs}\overline{n}_{q's'}(\overline{n}_{q"s"}+1)\delta(\omega_{qs}+\omega_{q's'}-\omega_{q"s"})\delta_{q+q'-q",G}, \tag{10a}$$

and

$$\overline{P}_{qs}^{q's',q"s"} = \frac{2\pi}{\hbar^2}\left|\Phi^{qs,q's',q"s"}\right|^2 \times (\overline{n}_{qs}+1)\overline{n}_{q's'}\overline{n}_{q"s"}\delta(\omega_{qs}-\omega_{q's'}-\omega_{q"s"})\delta_{q-q'-q",G}. \tag{10b}$$

The momentum selection rule is indicated by $\delta_{q\pm q'-q",G}$, denoting Kronecker delta function. The initial and final eigenstates and the anharmonic perturbation, $\Phi^{qs,q's',q"s"}$, of the crystal lattice Hamiltonian are required to determine the transition probability, which can be fixed by the aid of a force field [10,38]. By applying the second quantization transformation to the lattice Hamiltonian in atomic coordinates, we obtain an expression for the Hamiltonian in phonon coordinates. Following Srivastava's notation [10], the leading term of crystal anharmonicity is formulated as

$$\Phi_{anharm} = \frac{1}{3!}\sum_{\substack{qs,q's'\\q"s"}}\delta_{q+q'+q",G}\Phi^{qs,q's',q"s"}(a_{qs}^\dagger - a_{-qs})(a_{q's'}^\dagger - a_{-q's'})(a_{q"s"}^\dagger - a_{-q"s"}), \tag{11}$$

where

$$\Phi^{qs,q's',q"s"} = \frac{i}{\sqrt{N_o\Omega}}\sum_{\substack{b,b',b"\\\alpha,\alpha',\alpha"}}\left(\frac{\hbar^3}{8m_b m_{b'} m_{b"}\omega_{qs}\omega_{q's'}\omega_{q"s"}}\right)^{1/2}\Phi_{\alpha,\alpha',\alpha"}^{qb,q'b',q"b"}e_\alpha^{b,qs}e_{\alpha'}^{b',q's'}e_{\alpha"}^{b",q"s"}, \tag{11a}$$

and

$$\Phi_{\alpha,\alpha',\alpha''}^{qb,q'b',q''b''} = \sum_{l',l''} \left.\frac{\partial^3 \Phi}{\partial r_\alpha^{\binom{0}{b}} \partial r_{\alpha'}^{\binom{l'}{b'}} \partial r_{\alpha''}^{\binom{l''}{b''}}}\right|_0 e^{iq'\cdot l'} e^{iq''\cdot l''}. \tag{11b}$$

In the above, $l$ stands for the $l$th primitive unit cell of the crystal lattice, $b$ for the $b$th atomic basis in the primitive unit cell, $m_b$ for the mass of the atom residing at the $b$th basis, $N_o$ for the total number of the primitive unit cells, $\Omega$ for the primitive unit cell volume, $\alpha$ for one of the three orthogonal Cartesian directions in real space, $e_\alpha^{b,qs}$ for Cartesian component of phonon polarization vector, $a_{qs}^\dagger$ and $a_{-qs}$ for phonon ladder operators (phonon creation and annihilation operators), and $\Phi_{\alpha,\alpha',\alpha''}^{qb,q'b',q''b''}$ for the Fourier Cartesian component of the anharmonic (cubic) IFC; see Ref. [10] for more details.

### 2.3. Relaxation time approximation and models

As outlined in Sec. 1, the coupling terms in the collision kernel can be resolved through different treatments, mainly, relaxation time approximation, variational techniques, and iterative methods. Relaxation time approximation provides a phenomenological representation of the collision term that employs the relaxation time as a collective measure of phonon scattering rates. The relaxation time is thus the time scale for each of the excited phonon modes to relax exponentially to the equilibrium or steady state distribution. Phonon relaxation time is a function of phonon frequency, polarization, and temperature. Several techniques were used to derive an expression for phonon relaxation time, namely, time dependent perturbation theory, projection operator method, and double-time Green function method. All of these methods produce exactly the same expression for relaxation time, at least under SMRT approximation [10].

Under relaxation time approximation, the collision term is defined as

$$\left.\frac{\partial n_{qs}}{\partial t}\right|_{Collision} = -\frac{n_{qs} - \bar{n}_{qs}}{\tau_{qs}}. \tag{12}$$

This method can underestimate or overestimate the thermal conductivity [20], based on how relaxation times are obtained and the used solution scheme.

An expression for the effective intrinsic relaxation time can be derived from the perturbation theory-based collision term. Substituting Eq. (3) into Eq. (12), we get

$$-\left.\frac{\partial n_{qs}}{\partial t}\right|_{scatt} = \frac{n_{qs} - \bar{n}_{qs}}{\tau_{qs}} \simeq \frac{1}{k_B T} \frac{\psi_{qs} \bar{n}_{qs}(\bar{n}_{qs} + 1)}{\tau_{qs}}. \tag{13}$$

Comparing Eq. (13) with Eq. (9), the relaxation time, $\tau_{qs}$, for phonon mode $qs$ at temperature $T$ can be defined in terms of phonon collision operator and deviation from equilibrium distribution function as

$$\tau_{qs}^{-1} = \sum_{q's',q''s''} [\frac{\bar{P}_{qs,q's'}^{q''s''}}{\bar{n}_{qs}(\bar{n}_{qs}+1)} \frac{(\psi_{qs}+\psi_{q's'}-\psi_{q''s''})}{\psi_{qs}} + \frac{1}{2}\frac{\bar{P}_{qs}^{q's',q''s''}}{\bar{n}_{qs}(\bar{n}_{qs}+1)} \frac{(\psi_{qs}-\psi_{q's'}-\psi_{q''s''})}{\psi_{qs}}] \quad (14)$$

In this regard, different models are available with different levels of accuracy for representing the correlation between the relaxations of different modes (the coupling terms). The simplest is SMRT relaxation time. Under SMRT approximation (by letting $\psi_{q's'}=\psi_{q''s''}=0$), Eq. (14) reduces to

$$\tau_{qs}^{-1} \simeq \sum_{q's',q''s''} [\frac{\bar{P}_{qs,q's'}^{q''s''}}{\bar{n}_{qs}(\bar{n}_{qs}+1)} + \frac{1}{2}\frac{\bar{P}_{qs}^{q's',q''s''}}{\bar{n}_{qs}(\bar{n}_{qs}+1)}] \equiv \bar{P}_{3ph}. \quad (15)$$

SMRT approximation makes no distinction between normal and umklapp processes. It assumes that normal processes participate directly in restoring the equilibrium distribution in the same way as umklapp processes do. So, total relaxation time is simply given by Matthiessen's rule:

$$\frac{1}{\tau_{SMRT}} = \frac{1}{\tau_U} + \frac{1}{\tau_N}. \quad (16)$$

Callaway proposed an elaborate model that accounts for the non-resistive nature of normal processes by assigning a steady state distribution, for which quasi-momentum is an additional invariant of motion and takes coupling into consideration statistically. He rewrote the collision term under relaxation time approximation as [10,25]

$$\frac{\partial n_{qs}}{\partial t} = -\frac{n_{qs}-\bar{n}_{qs}}{\tau_r} - \frac{n_{qs}-n_{qs}(\mathbf{u})}{\tau_N}. \quad (17)$$

Where, $\tau_r$ is the relaxation time for the resistive processes (here, $\tau_r \equiv \tau_U$), and $\mathbf{u}$ is a unit vector in the same direction of the temperature gradient, i.e., $\mathbf{u} \parallel \nabla T$. From this, Callaway relaxation time $\tau_C$ can be formulated as [10]

$$\tau_C^{qs}(qs) = \tau_{SMRT}^{qs}(1+\frac{\beta}{\tau_N^{qs}}), \quad (18)$$

where

$$\beta = \frac{|\mathbf{q}|}{\omega^{qs} v_g^{qs}} \frac{\left\langle \omega^{qs} v_g^{qs} |\mathbf{q}| \frac{\tau^{qs}}{\tau_N^{qs}} \right\rangle}{\left\langle \frac{q^2}{\tau_N^{qs}} (1 - \frac{\tau^{qs}}{\tau_N^{qs}}) \right\rangle}, \tag{19}$$

and for any function $f$, $\langle f \rangle$ is its weighted BZ sum defined as

$$\langle f \rangle = \sum_{qs} f^{qs} \bar{n}_{qs} (\bar{n}_{qs} + 1). \tag{20}$$

The popularity of Callaway's model is attributed to its success to reproduce experimental thermal conductivity. Several modifications were proposed to improve its predictability, see the work of Allen and kinetic-collective model [39-41].

SMRT approximation eliminates the need to determine the deviation from equilibrium distribution functions, $\psi$, by ignoring the off-diagonal terms of phonon collision operator such that the unprimed deviation term in the denominator of Eq. (14) cancels out its analogue term in the numerator. However, this can introduce significant errors when the value of the off-diagonal elements is not negligible with respect to the diagonal elements, particularly at low temperature. On the other hand, Callaway's model and its counterparts considers the off-diagonal elements of the collision parameter solely for normal processes and in statistical average sense. To relax this approximation, Srivastava used linear response theory to derive a functional form for the off-diagonal elements in the case of umklapp processes [10]. He assumed that the deviation term is proportional to the magnitude of the projection of the wavevector on a unit vector, $\mathbf{u}$, in the direction of the applied temperature gradient, i.e., the deviational term was defined as [10]

$$\psi_{qs} \simeq \phi_q = \mathbf{q} \cdot \mathbf{u}. \tag{21}$$

Accordingly, it is dependent on the direction of the applied temperature gradient, but not on its magnitude. Consequently, thermal conductivity can show anisotropy, when isotropic continuum assumption is relaxed, which can have a significant impact on the value of thermal conductivity in different directions in the case of single crystals.

In spirit of Callaway's model, the expression Srivastava developed for the rate of change of phonon distribution function due to umklapp processes was formulated as

$$-\frac{\partial n_{qs}}{\partial t} \bigg|_U \simeq (P\phi)_{qs} = -\frac{\phi_q}{\tau_{U_{qs}}^{eff}} \bar{n}_{qs} (\bar{n}_{qs} + 1). \tag{22}$$

This yields different effective relaxation time for umklapp scattering, given by

$$\frac{1}{\tau_{U_{qs}}^{eff}} = \frac{(P\phi)_{qs}}{\phi_q \bar{n}_{qs}(\bar{n}_{qs}+1)} = \frac{\sum_{q's',q''s''}(\mathbf{G}\cdot\mathbf{u})(\bar{P}_{qs,q's'}^{q''s''} + \frac{1}{2}\bar{P}_{qs}^{q's',q''s''})}{(\mathbf{q}\cdot\mathbf{u})\bar{n}_{qs}(\bar{n}_{qs}+1)}. \tag{23}$$

Clearly this model does not affect normal processes, as $\mathbf{G}$ is equal to zero. The overall effective relaxation time developed in this model, called Srivastava's relaxation time ($\tau_S^{qs}$), takes the form

$$\tau_S^{qs} = \tau_{eff}^{qs}(1+\frac{\beta}{\tau_N^{qs}}), \tag{24}$$

where

$$\frac{1}{\tau_{eff}^{qs}} = \frac{1}{\tau_{U_{qs}}^{eff}} + \frac{1}{\tau_N^{qs}}. \tag{25}$$

Another way to account for the coupling terms can be achieved through seeking an iterative solution of BTE, which gives a more reliable prediction for the effective relaxation time, because it considers the function $\psi$ at individual mode level, without making apriori assumptions about its form. In this scheme, the iterative relaxation time is obtained from [3]

$$\tau_{qs}^{i+1} = \tau_{qs}^{SMRT}(1+\Delta_{qs}^i), \tag{26}$$

with

$$\Delta_{qs}^i = \sum_{q's',q''s''}[\frac{\bar{P}_{qs,q's'}^{q''s''}}{\bar{n}_{qs}(\bar{n}_{qs}+1)}\frac{(\tau_{q''s''}^i \upsilon_{g\|\nabla T}^{q''s''}\omega_{q''s''} - \tau_{q's'}^i \upsilon_{g\|\nabla T}^{q's'}\omega_{q's'})}{\upsilon_{g\|\nabla T}^{qs}\omega_{qs}} \\ + \frac{1}{2}\frac{\bar{P}_{qs}^{q's',q''s''}}{\bar{n}_{qs}(\bar{n}_{qs}+1)}\frac{(\tau_{q''s''}^i \upsilon_{g\|\nabla T}^{q''s''}\omega_{q''s''} + \tau_{q's'}^i \upsilon_{g\|\nabla T}^{q's'}\omega_{q's'})}{\upsilon_{g\|\nabla T}^{qs}\omega_{qs}}]. \tag{26a}$$

In this expression, $\upsilon_{g\|\nabla T}^{qs}$ is the component of the group velocity in the direction of the applied temperature gradient and $i$ indexes the iteration number. Like Srivastava's model, the iterative scheme can predict anisotropic thermal conductivity. The origins of this anisotropy are aggregated into the off-diagonal elements (coupling terms) of the phonon collision operator. In phonon kinetic theory, thermal conductivity can be calculated by plugging directly the desired effective relaxation time in a simple expression, as will be shown in Sec. 2.5. However, we need to emphasize that regardless of the model used to calculate the overall effective relaxation times, the distinction between the total mode-specific normal processes relaxation time ($\tau_N^{qs}$) and the total mode-specific umklapp processes relaxation time ($\tau_U^{qs}$) is important for the calculations.

## 2.4. Continuum approximation

Following the bottom-up approach, the coupling coefficients $\Phi^{qs,q's',q''s''}$ representing crystal anharmonicity and the harmonic eigenfrequencies both can be found from a classical or electronic-structure-based crystal potential using, for example, lattice dynamics [10]. The introduction of the continuum approximation further simplifies the relaxation time calculations in terms of average crystal properties. For example, Fourier components of phonon coupling constants are evaluated by fitting with a macroscopic Grüneisen parameters [10]. More approximations were introduced in literature regarding the dispersion curves, e.g., assuming isotropic continuum and/or linear dispersion relation, in addition to assuming a spherical BZ.

Srivastava [10] derived an expression for phonon relaxation time in continuum approximation in terms of Fourier component of phonon coupling constants. By applying long-wave approximation, he was able to reach a simple quantitative expression for the phonon coupling constants that depends only on one parameter, namely, mode-averaged (macroscopic) Grüneisen parameter, $\gamma$, along with the sound velocity, $\bar{\upsilon}_s$. Microscopic Grüneisen parameter is a thermodynamic property of the material and provides an average measure of crystal anharmonicity. The relaxation time under SMRT approximation, and similarly the equilibrium 3-phonon transition rates ($\bar{P}_{3ph}$), in Srivastava's notation are expressed as [10]

$$\tau_{qs}^{-1} = \frac{\pi \hbar \gamma^2}{\rho N_o \Omega \bar{\upsilon}_s^2} \sum_{q's'q''s''} \omega_{qs} \omega_{q_i's'} \omega_{q''s''} \times [\frac{\bar{n}_{q's'}(\bar{n}_{q''s''}+1)}{\bar{n}_{qs}+1} \delta(\omega_{qs}+\omega_{q's'}-\omega_{q''s''})\delta_{q+q'-q'',G} \\ + \frac{1}{2} \frac{\bar{n}_{q's'}\bar{n}_{q''s''}}{\bar{n}_{qs}} \delta(\omega_{qs}-\omega_{q's'}-\omega_{q''s''})\delta_{q-q'-q'',G}]. \quad (27)$$

As continuum approximation smears out all the structure, material density $\rho$ replaces the atomic masses in the above expression.

## 2.5. Phonon kinetic theory and computed properties

Phonon kinetic theory, treating phonon system as gas of bosons occupying the crystal lattice, can be used to derive an expression for lattice thermal conductivity of homogenous system by substituting directly in the linearized BTE under relaxation time approximation. At a given temperature, each normal mode contribution to the net heat flux can be measured in terms of the deviation of its phonon occupation number from the equilibrium distribution. In this case, the heat flux in an arbitrary direction with unit vector **n** is given by

$$q_{\bar{n}} = \frac{1}{V} \sum_{qs} \hbar \omega^{qs} (n^{qs} - \bar{n}^{qs})(\upsilon_g^{qs} \cdot \mathbf{n}). \tag{28}$$

But from Eq. (13) we have: $n^{qs} - \bar{n}^{qs} = \tau^{qs} c^{qs} (\upsilon_g^{qs} \cdot \nabla T)$, where $c^{qs}$ is the mode-specific contribution to the lattice specific heat. In harmonic approximation, with an average normal mode energy given by: $E_0^{qs}(T) = \hbar \omega_{qs} [\bar{n}_{qs}(T) + \frac{1}{2}]$, this yields

$$c^{qs} = \frac{\partial E_0^{qs}}{\partial T}\bigg|_{V,T} = \hbar \omega_{qs} \frac{\partial \bar{n}_{qs}}{\partial T} = \left(\frac{\hbar \omega_{qs}}{k_B T}\right)^2 \frac{k_B \exp(\frac{\hbar \omega_{qs}}{k_B T})}{\left(\exp(\frac{\hbar \omega_{qs}}{k_B T}) - 1\right)^2}. \tag{29}$$

Finally, recalling Fourier's equation:

$$q_{\bar{n}} = -(\mathbf{k}\nabla T) \cdot n, \tag{30}$$

where **k** labels the second-rank tensor of thermal conductivity. From this, an expression for the scalar lattice thermal conductivity, $k$, say in isotropic medium or cubic crystal system, can be directly derived

$$k = \sum_{qs} \frac{1}{3V} c^{qs} \tau^{qs} \upsilon_g^{qs} \cdot \upsilon_g^{qs}. \tag{31}$$

The harmonic component of the vibrational entropy, $S$, is another useful thermodynamic property defined as the first derivative of Helmholtz free energy, $F$, with respect to temperature under constant volume condition,

$$S = -\frac{\partial F}{\partial T}\bigg|_V = k_B \sum_{qs} [(\bar{n}_{qs} + 1) \ln(\bar{n}_{qs} + 1) - \bar{n}_{qs} \ln \bar{n}_{qs}]. \tag{32}$$

The 3-phonon phase space, $P_3$, is useful in enumerating the total number of scattering channels that satisfy energy conservation. To find that parameter, we can start by rewriting Fermi golden rule in the form [13]

$$P_i^f(3ph) = \frac{2\pi}{\hbar} |\langle f|H'|i\rangle|^2 \mathrm{D}(E_f - E_i). \tag{33}$$

In this equation D, known as joint Density of States (DOS), is the energy levels density of the final eigenstate (in continuum of energy levels). The total joint DOS for any mode, which is a measure of the number of decay channels, can be calculated from

$$D_s^{(\pm)}(q) = \sum_{s',s''} \int dq' \delta\left(\omega_s(q) \pm \omega_{s'}(q') - \omega_{s''}(q \pm q' - G)\right). \tag{34}$$

The $+$ ($-$) sign labels phonon creation (annihilation) events. The 3-phonon phase space is defined as the fraction of phonon phase-space available for 3-phonon scattering processes [42]. It is an integral quantity, averaged over all phonon modes, that is roughly inversely proportional to the thermal conductivity and is given by [42]

$$P_3 = \frac{2}{3S^3 V_{BZ}^2}(P_3^+ + \frac{1}{2}P_3^-), \tag{35}$$

where

$$P_3^{(\pm)} = \sum_s \int dq D_s^{(\pm)}(q). \tag{35a}$$

In this equation, S is the total number of polarization branches (for argon, S=3); the prefactor is used for the normalization, and the half factor in the fusion events term is used to avoid double summation.

### 3. Computational Approach

To quantitatively assess the effects of different approximations mentioned above, we calculate the intrinsic relaxation times using several models for dispersion relation, BZ shape and reciprocal lattice vector. For the effect of polarization type, the common assumption, under isotropic continuum approximation, is that eigenvectors are randomly oriented. Accordingly, no distinction between longitudinal modes and transverse modes are made based on the relative direction between the wavevector and polarization eigenvector. The values for lattice constant at 0 K (a= 0.53 nm) and the macroscopic Grüneisen parameter ($\gamma = 2.5$) were taken from Ref. [30]. The details of our numerical scheme are addressed in the sequel.

### 3.1. BZ sums and discretization

Point group symmetry properties of Face Centered Cubic (FCC) crystal structure is exploited to reduce the computation domain to only 1/48 portion of BZ, which is called Irreducible BZ (IBZ). We use simple cubic mesh for the discretization of IBZ, and applying one of the special k-points scheme to generate the grid points, which helps in getting a more efficient sampling of the total number of k points distributed uniformly over the sub-grid volume, cf. Ref. [23]. For FCC, the primitive cell volume ($\Omega$) is equal to one fourth of the conventional unit cell volume ($\Omega = \frac{a^3}{4}$). Moreover, the theoretical density is used in all simulations, for the sake of consistency.

There are two techniques to handle Dirac delta function in discrete summation, either to regularize it using an approximate discretized closed form of this delta function (e.g., rectangular function–unit pulse, suitable under narrow resonance approximation, is shown here), or to transform the discrete summation to a continuous integration. The expression for SMRT relaxation time, Eq. (27), after discretization using the two different methods yield, respectively

$$\tau_{qs}^{-1} = \frac{\pi \hbar \gamma^2}{\rho \Omega \bar{v}_s^2} \times \sum_{s's''} \sum_{i=1}^{M} w_i \frac{|q_i'|}{v_g^{q_i's'} \Delta \times (q_{ix}' + q_{iy}' + q_{iz}')} \omega_{qs} \omega_{q_i's'} \omega_{q''s''} \delta_{q+q_i'+q'',G}$$
$$\times [\frac{\bar{n}_{q_i's'}(\bar{n}_{q''s''}+1)}{\bar{n}_{qs}+1} \delta_{\omega_{qs}+\omega_{q_i's'}-\omega_{q''s''},0} + \frac{1}{2} \frac{\bar{n}_{q_i's'} \bar{n}_{q''s''}}{\bar{n}_{qs}} \delta_{\omega_{qs}-\omega_{q_i's'}-\omega_{q''s''},0}],$$
(36a)

$$\tau_{qs}^{-1} = 48 \times \frac{\hbar \gamma^2}{8\pi^2 \rho \bar{v}_s^2} \times \sum_{s's''} \sum_{i=1}^{M} \frac{S_{q_i'}}{v_{s'}} \omega_{qs} \omega_{q_i's'} \omega_{q''s''} \delta_{q+q_i'+q'',G}$$
$$\times [\frac{\bar{n}_{q_i's'}(\bar{n}_{q''s''}+1)}{\bar{n}_{qs}+1} \delta_{\omega_{qs}+\omega_{q_i's'}-\omega_{q''s''},0} + \frac{1}{2} \frac{\bar{n}_{q_i's'} \bar{n}_{q''s''}}{\bar{n}_{qs}} \delta_{\omega_{qs}-\omega_{q_i's'}-\omega_{q''s''},0}].$$
(36b)

Where, $S_{q_i'} = \Delta^2 (q_{ix}' + q_{iy}' + q_{iz}')/|q_i'|$. In these equations, $M$ is the total number of grid points, $\Delta$ is the mesh spacing, and $w_i$ is the fractional volume of the ith sub-grid with respect to the IBZ volume (more details on that are given in appendix A). In addition, $q_{ix}'$, $q_{iy}'$, and $q_{iz}'$ label the Cartesian components of $\mathbf{q}_i'$. Although the second method handles the delta function in an exact way in terms of group velocity, another difficulty arises due to the need to determine the surfaces of constant energy ($S_{q_i'}$), which usually introduces another source of approximation. When the surface of constant energy is hypothesized to be a plane normal to the wavevector, which is a common practice in most of the studies that followed this approach [20,24], the two expressions become equivalent. The regularization of the delta function using the truncated unit pulse is symbolized using an additional Kronecker delta function (in terms of phonon triplet frequencies). This symbol is used to indicate that energy conservation rule is enforced explicitly by filtering out all phonon triplets that do not meet the defined criteria for energy conservation, while using the same weight for all triplets passing the energy filter. This is in contrast to other extended representations using continuous distribution functions (for example, Gaussian or Lorentzian distribution functions) that treats energy conservation rules implicitly by accepting a weighted contribution from all phonon triplets, as will be explained in Sec. 3.3.

Consequently, we split our evaluation procedure into two steps. The first step is to find all candidate phonon triplets, for every single phonon mode in q space, that satisfy energy conservation and momentum selection rules. For this step, it is easy to show that the summation needs to run only over the BZ of the primed index (the q` sample points) without the need to go over the double primed index (as only one mode in the BZ of q`` space can satisfy the momentum selection rule for a given two modes in q and q` spaces, respectively). The second step is to evaluate the relaxation time for each individual phonon triplet, then add them together using Matthiessen's rule. In this regard, individual relaxation times of normal processes are added together and stored separately. The same is true for the umklapp processes. These last two distinct relaxation times, calculated for each mode, are added together using the suitable expression of one of the relaxation time models, described in Sec. 2.3, to find the overall effective relaxation time for each mode. Those calculated total relaxation times are then passed to the thermal conductivity calculation module to compute the bulk thermal conductivity. It is worth mentioning that a fixed-point iteration scheme was employed for the iterative relaxation time calculations, as indicated by Eq. (26). In addition, since phonon normal modes (wavevectors) are homogenously distributed over the BZ (under cyclic boundary conditions), joint DOS, after discretization, can be calculated using

$$D(\omega(q"s") \pm \omega(qs) - \omega(q's')) = \lim_{\Delta_q^3 (\equiv \Delta\omega) \to 0} \frac{\Delta_q^3(\omega(q"s") \pm \omega(qs) - \omega(q's'))}{V_{BZ}} \frac{N}{\Delta\omega}$$
$$\approx \frac{\Delta_q^3(\omega(q"s") \pm \omega(qs) - \omega(q's'))}{V_{BZ}} \frac{N}{\Delta\omega} = w_i \frac{N}{\Delta\omega}. \quad (37)$$

This expression is favorable for that it makes possible the exact point-wise calculation of 3-phonon phase space (at each phonon mode), instead of integration, and that within it, the connection between the delta function and the discretization scheme is defined clearly.

### 3.2. Dispersion models

To generate dispersion curves of solid argon, Mie-Lenard-Jones (6-12)—all neighbors interatomic potential with two fitting parameters, that reproduce the lattice constant and sublimation energy of argon at 0 K [30], was used in the present study. The dynamical matrix, $\mathbf{D(q)}$, for selected wavevectors was constructed from the Fourier transformed Cartesian-components of the second derivative of the crystal potential, evaluated at atoms equilibrium position. Subsequently, the three harmonic eigenvalues, for each wavevector, were found by solving the secular determinant of the diagonalized dynamical matrix. The secular determinant is given by

$$\left| \mathbf{D(q)} - \omega_{\mathbf{q}s}^2 \mathbf{I} \right| = 0, \quad (38)$$

where

$$\omega_{qs}^2 \mathbf{e}\begin{pmatrix}\mathbf{q}\\s\end{pmatrix} = \mathbf{D(q)e}\begin{pmatrix}\mathbf{q}\\s\end{pmatrix}, \tag{38a}$$

and

$$D_{3(b-1)+\alpha, 3(b'-1)+\alpha'}(\mathbf{q}) = \frac{e^{i\mathbf{q}\cdot(r_{(b')}^{(0)} - r_{(b)}^{(0)})}}{\sqrt{m^b m^{b'}}} \sum_{h'} \left.\frac{\partial^2 \Phi}{\partial r_\alpha^{(0)} \partial r_{\alpha'}^{(h')}}\right|_0 e^{i\mathbf{q}\cdot h'}. \tag{38b}$$

In the above, $\mathbf{I}$ is the identity matrix, and since argon has only one atom in the primitive cell ($b = 1$), the dynamical matrix size is $3\times 3$. Dispersion curves in the three high symmetry directions [001], [110], and [111], for the three different polarization types, were built by fitting the three eigenvalues of the selected wavevectors in each direction, using trigonometric functions, as shown in Fig. 1.

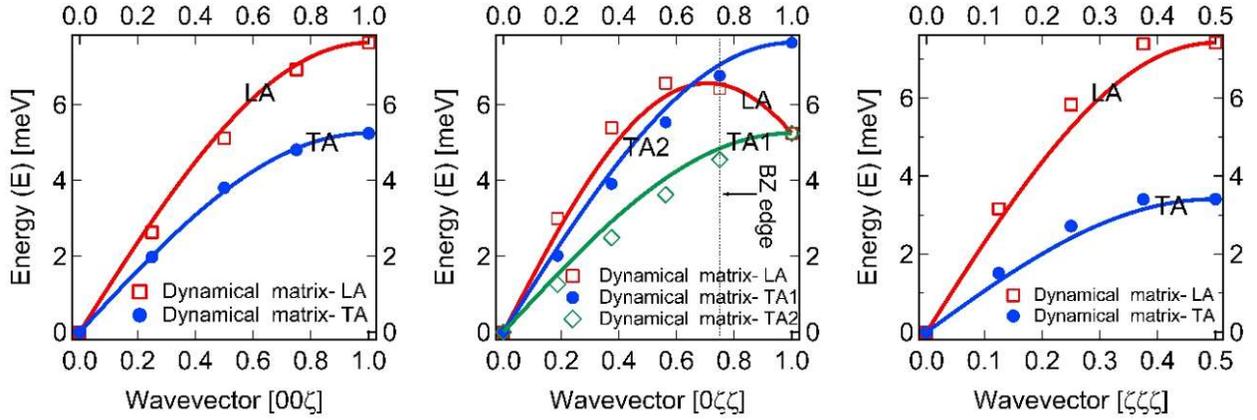

Fig. 1. Phonon dispersion curves (solid lines) in the high symmetry crystallographic directions [001], [110], and [111] for the three polarization branches: low energy Transverse Acoustic (TA1), high energy Transverse Acoustic (TA2), and Longitudinal Acoustic (LA); along with the eigenvalues for selected wavevectors calculated by solving the secular determinant of the dynamical matrix of argon, which was constructed using Mie-Lenard-Jones interatomic potential. $\zeta$ is the Cartesian component of wave vector in reciprocal lattice unit of $2\pi/a$, where a is the lattice constant. The two transverse branches are degenerate in the [001] and [111] directions.

Three dispersion models for ISOtropic continuum using only dispersion curves of one of the three high symmetry crystallographic directions are tested. They are named ISO1, ISO2, and ISO3, for [001], [110], and [111], respectively. In these models, when truncated octahedron BZ is used, the surfaces of constant energy are taken to be the boundaries of a set of concentric truncated octahedrons parallel to the external surface of the BZ. So, the frequency is assumed to be a function of one dimensionless parameter. A model for Semi-ANISOtropic medium (SANISO model), similar to the one used in [43] was examined as well. For this dispersion model, IBZ is divided (direction-wise) to three portions, each surrounds one of the principal high symmetry directions and angular isotropy is assumed within each portion. Each portion is represented by subtended solid angle extending from the origin to the external boundaries of IBZ in one of these three directions. We developed a Fully ANISOtropic model (FANISO), by interpolating between

Page | 17

the three high symmetry directions dispersion data, for a general point in the BZ. Thus, this model gives us the opportunity to directly incorporate the dispersion curves in the three high symmetry directions into an interpolation scheme over the reciprocal space to represent the cubic anisotropy without the need to solve the dynamical matrix for points in general directions (given its high computational cost and the requirement of adopting a force model or knowing the interatomic potential of the crystal). Practically speaking, the elimination of this step does not compromise the physics of the problem, as many of the force models used in lattice dynamic (for example, Born—von Karman force constants model, shell model, and bond-charge model) that reproduced successfully the measured phonon dispersions are not physically sound [10,11]. More elaborate interpolation scheme can be sought, as needed, to improve the agreement with experimentally measured phonon DOS, however the main goal here is to elucidate the impact of phonon DOS on the phonon collision operator as a result of the size of uncertainty associated with normal modes frequencies in the low symmetry directions, a procedure that is usually ignored in the assessment of the calculated dispersion curves. It is important to keep in mind that within our quasi-continuum approximation, we assumed the effect of eigenvector directions is cancelled out. In addition, group velocities are postulated to be in the same direction as the wavevectors, which sounds appropriate when dealing with cubic system with the highest degree of symmetry among crystal systems.

For the fitting scheme, we start by defining a dimensionless parameter ($r$) for each wavevector as the ratio between its magnitude and the length of the wavevector pointing in the same direction with its extremum lies on the BZ surface ($\mathbf{q}|_{BZS}$), thus, $0 \leq r \leq 1$. Accordingly, for any given wavevector $\mathbf{q}$ with its polar and azimuthal angle in spherical coordinate system denoted by $\theta$ and $\phi$, respectively, its associated dimensionless parameter is calculated from

$$r(\mathbf{q}(\theta,\phi)) = \frac{|\mathbf{q}(\theta,\phi)|}{|\mathbf{q}|_{BZS}(\theta,\phi)|}. \tag{39}$$

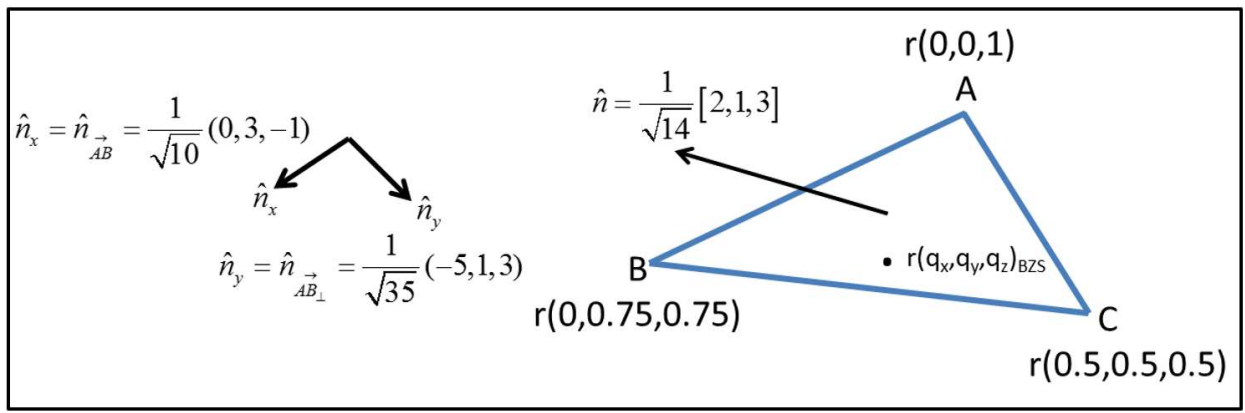

Fig. 2. Mapping scheme of BZ sample points on triangular surfaces.

This dimensionless parameter is calculated and stored with the three Cartesian components of the wavevector ($q_x, q_y, q_z$) for each sample point. In addition, this parameter is used to map the sample point onto one of the reduced triangles depicted schematically in Fig. 2. For any reduced triangle, its vertices, namely, A, B, and C belongs, respectively, to [001], [011], and [111] crystallographic directions. Moreover, all the points on any reduced triangle share the same normalized dimensionless parameter $r$. When FANISO model is invoked, two additional coordinates (for the reduced 2D representation of Fig. 2) are precomputed using

$$x(\mathbf{q}) = (\mathbf{q} - \mathbf{A}) \cdot \hat{n}_x,$$
$$y(\mathbf{q}) = |(\mathbf{q} - \mathbf{A}) \times \hat{n}| \cdot \hat{n}_y. \qquad (40)$$

The dispersion curves in the three high symmetry directions for different polarization branches ($\omega_{\mathbf{q}s}^{hs}$) are generated by fitting the dispersion data points calculated by solving the dynamical matrix using trigonometric functions ($f_i^{hs}(r, s)$), where the $i$ index labels the direction, i.e.,

$$\omega_{\mathbf{q}s}^{hs} = f_i^{hs}(r, s). \qquad (41)$$

The functional form used in the fitting procedure here is obtained from

$$f_i^{hs}(r, s) = f_{0,i}^{s} + A_i^{s} \sin(B_i^{s} r + \varphi_i^{s}). \qquad (42)$$

In Eq. (42), $f_{0,i}^{s}, A_i^{s}, B_i^{s}$, and $\varphi_i^{s}$ are the fitting parameters, which are constrained by the boundary conditions given by

$$f_i^{hs}(r = 0, s) = 0,$$
$$\upsilon_{g,i}^{hs}(r = 1, s) \equiv \frac{\partial f_i^{hs}(0, s)}{\partial r} = 0, \text{ for i = [001], or [111]}. \qquad (43)$$

For ISO1, ISO2, and ISO3 models, only dispersion curves in the corresponding direction are used. While for SANISO model, the sample point in general direction is assigned the same values associated to the nearest vertex of the reduced triangle to which it belongs. Finally, FANISO model uses the linear interpolation scheme governed by

$$\omega_{\mathbf{q}s} = \sum_{i=1}^{3} f_i^{hs}(r, s) w_i(x, y). \qquad (44)$$

Where, $w_i(x,y)$ is the triangular element shape function in its standard form for linear three-node triangle:

$$w_i(x,y) = \frac{1}{2\Delta}(a_i + b_i x + c_i y), \qquad (45)$$

and $\Delta$ is the area of the rectangular:

$$\Delta = \frac{1}{2}\begin{vmatrix} 1 & x_A & y_A \\ 1 & x_B & y_B \\ 1 & x_C & y_C \end{vmatrix} = \frac{1}{2}(b_A c_B - b_B c_A). \qquad (46)$$

In addition, $a_i$, $b_i$, and $c_i$ are interpolation coefficients, and can be determined by solving

$$\begin{bmatrix} a_1 & b_1 & c_1 \\ a_2 & b_2 & c_2 \\ a_3 & b_3 & c_3 \end{bmatrix} = \begin{bmatrix} (x_B y_C - x_C y_B) & (y_B - y_C) & (x_C - x_B) \\ (x_C y_A - x_A y_C) & (y_C - y_A) & (x_A - x_C) \\ (x_A y_B - x_B y_A) & (y_A - y_B) & (x_B - x_A) \end{bmatrix}. \qquad (47)$$

Using this trigonometric function fitting scheme, phonon dispersion curves and group velocities in the three high symmetry directions are generated and depicted in Figs. 1 and 3, respectively.

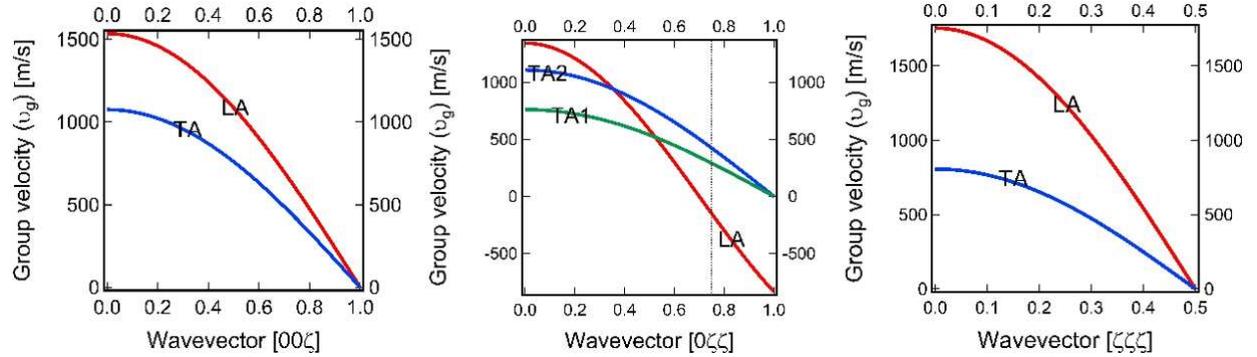

Fig. 3. Phonon group velocities in the three high symmetry directions [001], [110], and [111] for the different polarization branches, calculated by the direct differentiation of the dispersion curves.

### 3.3. Conservation rules and Dirac delta function handling

The search module of the code for candidate phonon triplets checks for the two conservation rules. Searching over phonon triplets that satisfy momentum selection rule is done separately at the beginning. This is carried out by looping over all mesh-points and polarization branches in q domain. An internal loop running over all mesh-points and polarization branches in q` domain is implemented. This includes a nested loop for each polarization branch in q`` domain (but not the sample points), which works on finding the phonon wavevector that satisfy momentum selection rule (for a given two wavevectors in q and q` space) for each of fusion and fission processes, respectively, according to

$$\begin{array}{c} \textit{Normal process} \\ \begin{cases} q``_x = q_x + q`_x \\ q``_y = q_y + q`_y \\ q``_z = q_z + q`_z \end{cases} \end{array} \begin{array}{c} \rightarrow \textit{fusion} \leftarrow \\ \omega + \omega` \rightarrow \omega`` \end{array} \begin{array}{c} \textit{Umklapp process} \\ \begin{cases} q``_x = q_x + q`_x - G_x \\ q``_y = q_y + q`_y - G_y \\ q``_z = q_z + q`_z - G_z \end{cases} \end{array}, \qquad (48)$$

$$\begin{cases} q``_x = q_x - q`_x \\ q``_y = q_y - q`_y \\ q``_z = q_z - q`_z \end{cases} \begin{array}{c} \rightarrow \textit{fission} \leftarrow \\ \omega \rightarrow \omega` + \omega`` \end{array} \begin{cases} q``_x = q_x - q`_x + G_x \\ q``_y = q_y - q`_y + G_y \\ q``_z = q_z - q`_z + G_z \end{cases}. \qquad (49)$$

Although the energies of any crystallographically equivalent directions are the same (as they depend only on the dimensionless parameter) such that they can be represented by IBZ, the momentum summation is a vectorial type. Accordingly, due to the momentum selection rule, we must consider the whole BZ in the q` space (not only the IBZ). This is done, for any sample point in q` space, by looping over all different indicial permutations of the three components of its Cartesian coordinates (including the distinction between the positive and negative values of each component). If the resultant wavevector in q`` domain lies out the first BZ, then it is recorded as un umklapp process and the suitable reciprocal wavevector ($G$) is used to bring it back to the BZ and the new reduced vector is used instead. Otherwise, it is registered as a normal process. In this regard, it is worth to mention that G in this case for FCC is restricted to $\frac{2\pi}{a}<002>$, $\frac{2\pi}{a}<111>$, and $\frac{2\pi}{a}<022>$ families of crystallographic directions.

At each search iteration, the phonon triplet that satisfies momentum selection rule is first picked then checked whether it satisfies energy conservation rule as well. This starts by looping over different polarization branches to determine the energy of the phonon mode corresponding to the calculated wavevector in q`` space that satisfy the momentum conservation rule. Based on whether it is fission or fusion event, the suitable expression is used, and two energies are calculated in each iteration over the double primed polarization loop (s``). This is followed by the explicit filtration using the discrete numerical approximation of the Dirac delta function. If the energy difference is less than a certain limit (determined by the definition used for the regularized delta function, see next), the phonon triplet is accepted as a candidate for 3-phonon processes, otherwise the triplet is disregarded. SMRT relaxation time for each phonon triplet passed conservation rules is calculated for fusion and fission events, respectively, using

$$\tau^{-1}(\omega_{qs} + \omega_{q_i's'} \to \omega_{q"s"}) = \frac{\pi\hbar\gamma^2}{\rho\Omega\upsilon_s^2} \times \frac{w_i}{|\Delta\omega_i'|} \omega_{qs}\omega_{q_i's'}\omega_{q"s"} \times \left(\frac{\overline{n}_{q's'}(\overline{n}_{q"s"}+1)}{\overline{n}_{qs}+1}\right), \tag{50}$$

$$\tau^{-1}(\omega_{qs} \to \omega_{q_i's'} + \omega_{q"s"}) = \frac{1}{2}\left[\frac{\pi\hbar\gamma^2}{\rho\Omega\upsilon_s^2} \times \frac{w_i}{|\Delta\omega_i'|} \omega_{qs}\omega_{q_i's'}\omega_{q"s"} \times \left(\frac{\overline{n}_{q's'}\overline{n}_{q"s"}}{\overline{n}_{qs}}\right)\right]. \tag{51}$$

In our implementation, we followed the regularization approach of the Dirac delta function, refer to the discussion in Sec.3.1. For this purpose, three different mathematical representations were tested, namely, unit pulse, Lorentzian and Gaussian function. The general definition of the unit pulse (rectangular) function, under narrow resonance approximation, can be formulated using the Heaviside step function. Our approximate form yields

$$\delta(x-x_0) = \begin{cases} \lim_{\varepsilon\to 0}\frac{1}{\varepsilon}, & |x-x_0| \leq \frac{\varepsilon}{2} \\ 0, Otherwise \end{cases} \approx \begin{cases} \frac{1}{\Delta_i}, & |x-x_0| \leq \frac{\Delta_i}{2} \\ 0, Otherwise \end{cases} \equiv \frac{1}{\Delta_i}\delta_{x_i,x_0}. \tag{52}$$

Apparently, the common property between this approximate function and the original one is that both integrate to unity. By close examination of the correspondence between the regularization approach and the second approach mentioned at the beginning of this section (by carrying out the delta function integration first), it can be observed that the width $\Delta$, and hence the height, of the rectangular function is not arbitrary. This is very important, as this affects the number of phonon triplets that can pass the energy conservation test. When the width is selected to be large, many triplets will pass the test and energy conservation rule will be violated. On the other hand, when the width is so small almost no phonon triplet will pass the test. Unlike many other studies that used the width as an adjustable parameter, we here fixed it on the basis of the aforementioned correspondence argument.

In terms of the BZ sample points, the unit pulse in Eq. (52) can have three possible different forms:

$$\delta(\omega_{qs}+\omega_{q_i's'}-\omega_{q"s"}) = \begin{cases} \frac{1}{|\Delta\omega_i'|}, & |\omega_{qs}+\omega_{q_i's'}-\omega_{q"s"}| \leq \frac{|\Delta\omega_i'|}{2} \\ 0, Otherwise \end{cases} \to \tau(q,s,\omega), \tag{53a}$$

$$\delta(\omega_{qs}+\omega_{q_i's'}-\omega_{q"s"}) = \begin{cases} \frac{1}{\upsilon_g^{\omega_i's'}|\Delta q_i'|}, & |\omega_{qs}+\omega_{q_i's'}-\omega_{q"s"}| \leq \frac{\upsilon_g^{\omega_i's'}|\Delta q_i'|}{2} \\ 0, Otherwise \end{cases} \to \tau(s,q(\omega)), \tag{53b}$$

$$\delta(\omega_{qs} + \omega_{q_i's'} - \omega_{q"s"}) = \begin{cases} \dfrac{1}{\upsilon_g^{q_i's'}|\Delta q_i'|}, & |\omega_{qs} + \omega_{q_i's'} - \omega_{q"s"}| \leq \dfrac{|\upsilon_g^{q_i's'}|\Delta q_i'|}{2} \\ 0, & Otherwise \end{cases} \to \tau(s,\omega(q)). \quad (53c)$$

Having three different forms is attributed to the dispersive relation and whether we treat the dependency of the relaxation time on the wavevector and frequency explicitly or implicitly, i.e., based on whether we used the group velocity of the sample point (implicit representation of energy dependence) or the energy difference between the furthest two surface of constant energy that belongs to the sub-grid (explicit representation of energy difference) to determine the energy spacing. The third version arises when we treat the problem explicitly in energy domain instead of q space (the wavevector of the sample point typically located in the center of the frequency bin is calculated using invertible dispersion function, and the associated volume element in q space is usually assumed to be spherical). Although many studies used this treatment (in particular those who transformed the BZ sums to frequency domain integral), realistic phonon spectrum results in phonon DOS with several Van Hove singularity points in frequency domain [23]. Accordingly, a very fine mesh (in frequency domain) is always needed to deal with this difficulty and the convergence is slow. Consequently, we abandoned this method. It should be mentioned that the consistency requirements dictate the use of the same definition for the energy conservation test and the height of the function.

For the regions where the dispersion relation is linear, the frequency bin spacing for a given sub-grid in q space (remember that the mesh is irregular in the frequency domain) can be given directly by the group velocity of the sample point (taken at the centroid of the sub-grid) times the radial spacing in the q space and the following relation holds true:

$$\delta(\omega_{qs} + \omega_{q_i's'} - \omega_{q"s"}) \approx \dfrac{\upsilon_g^{q_i's'}|\Delta q_i'|}{|\Delta \omega_i'|} \delta(\omega_{qs} + \omega(q_i',s) - \omega_{q"s"}). \quad (54)$$

However, a correction term should be used to account for the finite number of q points used in the sampling scheme of BZ: $\tau(q,s,\omega) = \dfrac{\upsilon_g^{qs}|\Delta q|}{|\Delta \omega|}\tau(s,\omega(q))$. Where, $\upsilon_g^{qs} = \lim\limits_{\Delta q \to 0} \dfrac{\Delta \omega}{\Delta q} \equiv \dfrac{d\omega}{d\mathbf{q}}$. The different versions of the approximate function were tested and no remarkable difference on the results were observed. We preferred to adhere to Eq. (53a) for the sake of consistency, as the energy conservation test is done separately and we believed it is more appropriate to deal with this part in the energy domain rather than q space. Accordingly, the relaxation time for each individual event was finally calculated using Eqs. (50) and (51) for fusion and fission, respectively. Again, the half factor in the expression used for fission event was put to avoid double summation.

In addition to the truncated uniform representation, extended representation of Dirac delta function was also tested using Gaussian distribution function

$$\delta(\omega_i - \omega_0) = \frac{1}{\sqrt{2\pi}\sigma} e^{\frac{(\omega_i - \omega_0)^2}{2\sigma^2}}, \qquad \sigma = \frac{\Delta_{\omega_i}}{\sqrt{2\pi}}, \tag{55}$$

and Lorentzian distribution function

$$\delta(\omega_i - \omega_0) = \frac{1}{\pi} \frac{\varepsilon}{(\omega_i - \omega_0)^2 + \varepsilon^2}, \qquad \varepsilon = \frac{\Delta_{\omega_i}}{\pi}. \tag{56}$$

The smearing factor (the standard deviation) of the distributions was fixed by making a connection to the energy bin spacing (which is mesh density dependent) based on the correspondence argument. The criterion was to get the same maximum amplitude (height), located at the center ($\omega_i = \omega_0$), of the Dirac function for the different mathematical representation used. By doing this, we obtained results in fair agreement (up to a constant) with the ones we get from the rectangular function at high temperatures. However, as we showed in previous studies [44-46], the use of Lorentz distribution to represent Dirac delta, based on physical argument, is critical at low temperature to capture the peak thermal conductivity and achieve good agreement with experiment. When the extended definition is preset, the energy conservation filter is turned off, which makes the calculations computationally much more expensive. However, their use can serve also as a benchmarking for the computationally cheaper truncated definition

## 4. Results and Discussions

In this section, we test the different approximations considered in present study by computing the intrinsic lattice thermal conductivity, spectral thermal conductivity, and phonon properties of FCC argon. This includes phonon DOS, 3-phonon scattering phase space, relaxation times, Debye characteristic temperature, harmonic isochoric specific heat, and vibrational entropy. In addition, we present the convergence analysis of our model used in benchmarking these approximations. Different approximations and models considered in the present study are summarized in Table 1. These approximations include dispersion models (ISO1, ISO2, ISO3 SANISO, and FANISO), dispersion relations (linear versus nonlinear), BZ shape (truncated octahedron versus sphere), relaxation time models (SMRT, Callaway, Srivastava, and the iterative scheme), as well as different mathematical representations of Dirac delta function that was discussed in Sec. 3.3.

Table 1. Summary of different models and approximations considered in the present study.

| Dispersion models | | Dispersion relation | BZ structure | Relaxation time models | Dirac delta function treatment |
|---|---|---|---|---|---|
| Isotropic Continuum | ISO1 [001] | Linear $\omega(\mathbf{q},s) \propto |\mathbf{q}|$ | Sphere $|\mathbf{G}| = \dfrac{4\pi}{a}, \ \mathbf{G} \parallel \mathbf{q}``$ | SMRT | Unit pulse |
| | ISO2 [110] | | | Callaway | Lorentzian |
| | ISO3 [111] | Nonlinear | | Iterative | Truncated Lorentzian |
| SANISO Semi-ANISOtropic | | $\omega(\mathbf{q},s) \propto \sin(A|\mathbf{q}|+\varphi)$ $A$ and $\varphi$ are fitting parameters. | OC (truncated octahedron) $\mathbf{G} \in \dfrac{2\pi}{a}\{\langle 002 \rangle, \langle 220 \rangle, \langle 111 \rangle\}$ | Anisotropic thermal conductivity Srivastava | Gaussian |
| FANISO Fully ANISOtropic | | | | | Truncated Gaussian |

Table 2. Calculated 3-phonon scattering phase space (P$_3$), Debye characteristic temperature ($\theta_D$)*, and mode-averaged root-mean-square group velocity (<υ$_g^2$>$^{1/2}$) for different dispersion models and BZ structures (sphere versus truncated octahedron, OC) for both linear and nonlinear dispersion relations.

| Dispersion Model | P$_3$ | | $\theta_D$ (K) | | <υ$_g^2$>$^{1/2}$ (m/s) | |
|---|---|---|---|---|---|---|
| | linear | nonlinear | linear | nonlinear | linear | nonlinear |
| ISO1/Sphere | 0.0034 | 0.0014 | 71.65 | 82.44 | 1184 | 558 |
| ISO1/OC | 0.0039 | 0.0013 | 71.31 | 82.27 | 1166 | 551 |
| ISO2/Sphere | 0.0035 | 0.0017 | 72.17 | 82.08 | 983 | 586 |
| ISO2/OC | 0.0034 | 0.0016 | 71.93 | 81.17 | 968 | 579 |
| ISO3/Sphere | 0.0036 | 0.0016 | 59.29 | 69.43 | 939 | 535 |
| ISO3/OC | 0.0043 | 0.0014 | 59.15 | 68.85 | 925 | 529 |
| SANISO/OC | 0.0044 | 0.0023 | 67.33 | 76.87 | 1005 | 557 |
| FANISO/OC | 0.0167 | 0.0111 | 67.83 | 77.59 | 1005 | 545 |

* $\theta_D$ ($\equiv \theta_\infty$) is the high temperature limit value evaluated from the second moment of the phonon frequency spectrum ($\mu_2 = \dfrac{\int \omega^2 D(\omega) d\omega}{\int D(\omega) d\omega}$), where $\theta_D = \hbar \sqrt{\dfrac{5\mu_2}{3}}/k_B$ [47]

Table 2 provides the calculated three-phonon phase space (P$_3$), the high temperature limit of Debye characteristic temperature, and the mode-averaged (DOS weighted) root-mean-square group velocity (<υ$_g^2$>$^{1/2}$) using different dispersion models. For each dispersion model, the parameters calculated using nonlinear dispersion relation are compared with the same dispersion model when linear dispersion relation is assumed. It is worth to mention here that linear dispersion was established using the frequency values at the BZ center and boundaries, for different orientations and polarization branches, not the sound

velocity—defined as the group velocity at Γ point. In addition, values for spherical BZ approximation are also reported in the case of isotropic dispersion models for the sake of comparison with the actual truncated octahedral BZ. From Table 2, we can infer several remarks about the interplay between different approximations used. By checking the estimated values of integral quantities listed in Table 2, which appear in macroscopic expressions for the thermal conductivity, we realize that the combined effect of adopting these simplifications can result in overestimation or underestimation of the calculated thermal conductivity. This is mainly because the relative errors, due to the introduction of such approximations, in the evaluation of different parameters, constituting thermal conductivity expression, goes in the two opposite directions. For example, while linear dispersion approximation increases the value of 3-phonon scattering phase space (which means higher phonons scattering rates and smaller relaxation times) and produces smaller value for Debye characteristic temperature, it yields higher values for the average group velocity. Accordingly, the overall impact of this approximation on the thermal conductivity cannot directly be inferred. This elucidates the role played by error cancellation for simple models, which ignore real phonon states structure, to reproduce experiment at certain temperature range. However, spectral properties should always be examined for a model to be considered reliable.

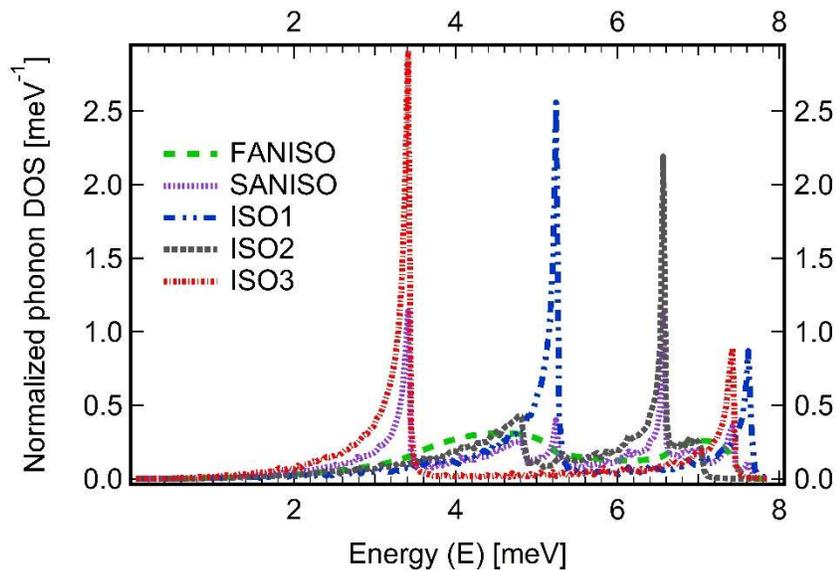

Fig. 4. Phonon density of states for different dispersion models using nonlinear dispersion relations.

Fig. 4 shows the simulated phonon DOS for different dispersion models. It can be seen in this figure that isotropic continuum models exaggerate the height of the resolved peaks, while underestimate the phonon DOS for the rest of the spectrum. The same behavior was reported by Houston [48] in his method to calculate the phonon spectrum by the use of Kubic harmonics expansion (basis functions derived from spherical harmonics and symmetric under cubic point group operations) using dispersion curves in the three high symmetry directions. This behavior is now well understood to be a direct result of the existing points with Van Hove singularity near (or at) the BZ surface in these directions. Under isotropic dispersion assumption and for a given polarization branch, all of these points lie on the same constant-energy-surface. On the other hand, FANISO and, to some extent, SANISO models alleviate this behavior and resemble better agreement with spectrum calculated from the lattice dynamic approach (by solving the secular determinant for many points in general directions and applying suitable interpolation technique to get the values for the q-points sampled on finer BZ mesh).

### 4.1. Convergence analysis and Dirac delta function

In the relaxation time calculation, the computational domain is the IBZ with a fixed size and the variable parameter is the mesh density (number of sample points). Accordingly, convergence study is usually performed by plotting the simulated thermal conductivity as a function of the number of mesh points. Applying this method, we found that, in respect of the mesh density, thermal conductivity converges fast in our scheme (setting $L_p$ to a value of 17, corresponding to a total number of 545 mesh points in IBZ and 26160 sample points in the whole BZ, was typically sufficient for temperatures above peak thermal conductivity temperature of 8 K). In comparison with other works, only 7 sample points along [001] crystallographic directions were enough to get a good estimation of thermal conductivity at 20 K, without the need to use additional finer gridding for a mesh with 9 sample points along [001] directions as was needed in Ref. [20]. In addition, our scheme outperforms the algorithm presented by Turney et al. [15] who had to use extrapolation method (by plotting the inverse of the thermal conductivity versus the inverse of the mesh density) to get a value for thermal conductivity corresponding to the highly dense mesh. He justified this step as the way to account for the contribution of phonon modes in the long-wavelength limit (located near the BZ center), since the sub-grid enclosing the BZ center is usually ignored in the calculation (point of singularity). Chernatynskiy et al. [20] refuted this argument and attributed this behavior to the inaccurate calculation of the area of surfaces of constant energy and the frequent occurrence of situations where the delta function root does not lie on a regular mesh point. Since our algorithm overcomes these difficulties we can assert confidently that the additional extrapolation step proposed by Turney et al. is not necessary in general. In addition, one of our observation from studying the convergence behavior was that 3-phonon phase space ($P_3$), with respect to thermal conductivity, is way sensitive to the convergence limit of calculated relaxation times. So, we recommend it as a better convergence criterion, particularly when spectral properties are under consideration.

One of the goals of this study is to investigate the origin of oscillations in relaxation time profile, which was reported in previous studies under continuum of energy levels approximation, to emphasize whether this behavior is authentic or spurious (more details on that will be provided in Sec. 4.2). We needed first to determine the numerical scheme sources of error to single them out. The spurious oscillations in relaxation time profile can be ascribed to three main parameters, namely, the number of BZ sample wavevectors (mesh density), the choice of Dirac delta function width/height versus mesh spacing (~ number of scattering channels), and Dirac delta function shape. Several measures have been applied to improve the convergence including: a) switching to the mesh spacing in frequency domain for delta function definition, b) testing the impact of defining the frequency at each sub-grid as the value at the centroid of the sub-mesh in frequency domain (irregular mesh) rather than the frequency of the q point lying in the center of the element of volume in q space (uniform mesh), and c) switching the summation in q` domain from the regular mesh points to the center of the delta function. For the elimination of round off errors, the calculations switched from frequency domain to energy domain. In addition, to further improve the convergence, the order of the mathematical operations in the relaxation time expression implemented in the computer code was rearranged to operate on the parameters of the same order of magnitude firstly and the wavevectors were defined in reduced reciprocal unit length (i.e., normalized by a factor of $2\pi/a$).

To investigate the impact of the mesh density and the number of iterations on the simulation results, a convergence study was conducted. It turned out that the convergence pattern exhibits temperature dependence. Figure 5(a) demonstrates SMRT conductivity for different mesh density. Apparently, dense meshes are needed at low temperature, while at high temperature coarse meshes are fairly sufficient. Again, this is attributed to the significance of the contribution of low energy phonons at low temperature, so very dense mesh is needed to sample enough wavevectors in that limit. At high temperature (> 20 K), it is obvious that conductivity is largely insensitive to the details of phonon states, which justify the success of theoretical models that depends on one effective macroscopic measure of the frequency spectrum (Debye frequency) at this limit, e.g., Slack model [49]. The high computational cost of the iterative scheme and the implicit representation of energy conservation forced us to limit the number of iterations to 4 in all simulations. Figure 5(b) ascertains the adequacy of this number. The inset gives the relative error with maximum that did not exceed 3% (about 20K). The three different mathematical representations of Dirac delta, considered here, help in providing insight on the impact of the approximation that three-phonon processes are of elastic type, by assessing different levels of strictness in applying energy conservation rule. Figure 5(c) depicts thermal conductivity profile as a function of temperature for the three cases using very coarse mesh (with $L_p$ set to a value of 51, corresponding to a total number of 12401 mesh points in IBZ and 595248 sample points in the whole BZ). Obviously, the three profiles behavior at high temperatures is the same, with Gaussian distribution curve coincide with Lorentzian distribution curve in this regime. However, in the low temperature regime, Lorentzian distribution has a unique impact on the profile introducing a characteristic peak. This signature is a direct result of the skewness of Lorentzian function and is persistent for different mesh density with a peak at almost the same location, as noted from Fig. 5(a). Loose convergence criterion affects mainly the calculated values for thermal conductivity below 3 K, without affecting the profile behavior versus temperature. On the other hand, for a very coarse mesh (representing the case of poor Brillouin zone sampling, and diverging results) a peak can be obtained at completely different temperature (~ 1-2 K), when unit pulse representation of the Dirac delta function is invoked, as displayed in Fig. 5(d). As in our definition unit pulse representation gives broader peak, compared to Gaussian and Lorentzian function, this representation underestimates the conductivity at high temperature by a factor of half, see Fig. 5(c). This conforms with the value of $P_3 = 0.02$ in the case of unit pulse, which is twice the same parameter in the case of Lorentzian and Gaussian distributions (with $P_3 = 0.01$ for both). However, this is not the case in the low temperature regime, as the interactions between phonon triplets that strictly satisfy energy conservation are not the most important. In this regard, it is worth to mention that truncated version of Gaussian and Lorentzian distributions was tested, where the magnitude of these functions are set to zero for energies with absolute difference from the root larger than one standard deviation. This is similar to the explicit energy conservation representation followed in the unit pulse function representation, where energy filter was set, see Sec.3.3. These truncated representations produced higher magnitudes for thermal conductivity, due to constraining three-phonon phase space, which made their thermal conductivity prediction, under SMRT approximation, comparable to the iterative scheme prediction, when the extended distribution is invoked. At the same time, the characteristic peak of the Lorentzian was lost when the truncated representation was in effect, which confirms the role played by the heavy tail of Lorentzian function at low temperature.

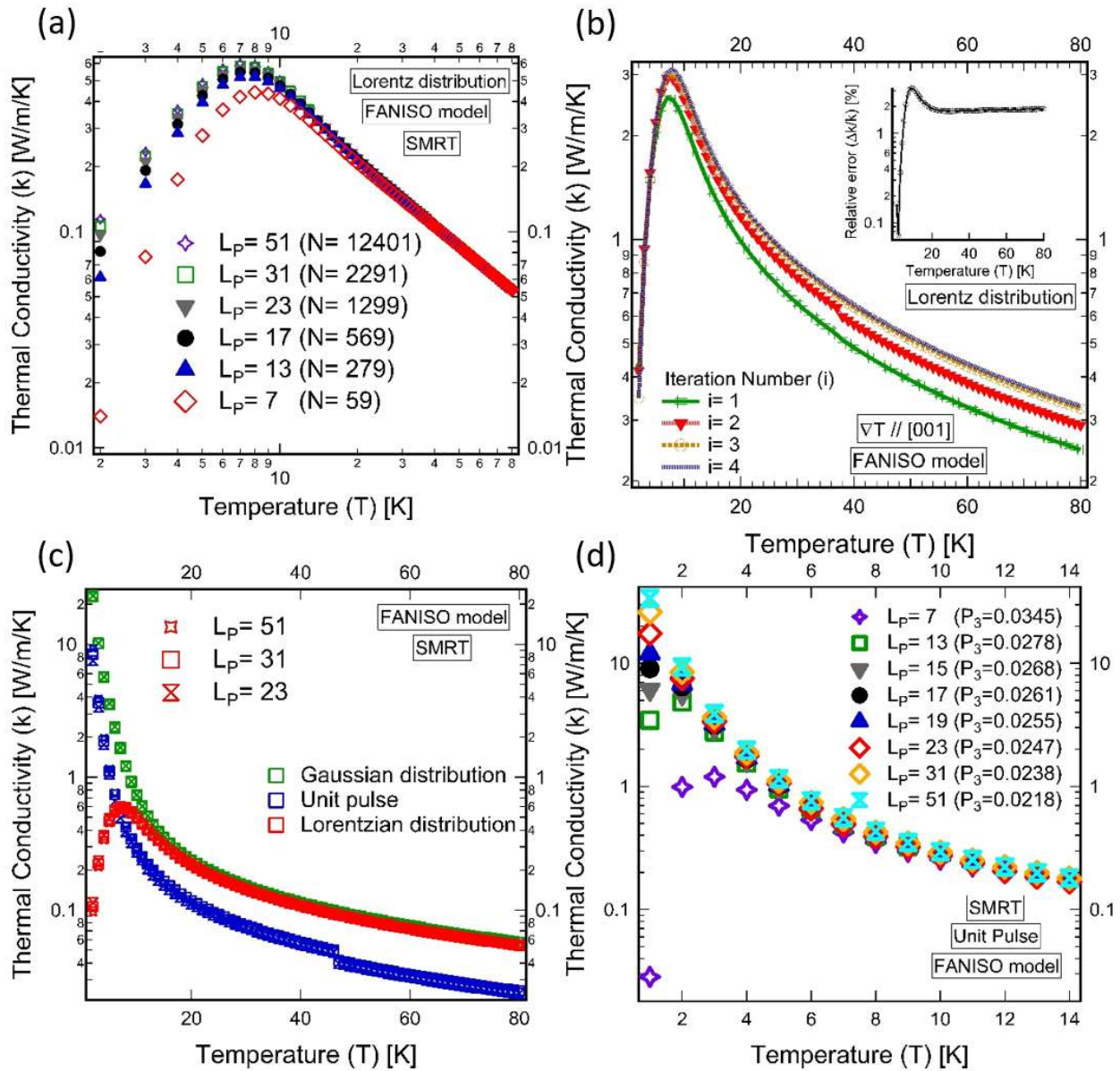

Fig. 5. (a) Convergence of thermal conductivity profile using several mesh densities based on SMRT approximation, FANISO dispersion model, and Lorentz distribution (N is the total number of wavevectors used to sample the irreducible BZ, and $L_P$ is the number of sample points along [001] crystallographic direction. (b) Thermal conductivity profile (as a function of temperature) for the first four iterations of the iterative solution for the same model, using $L_P$=23, starting with the SMRT solution for $i$=0. The inset reports the absolute relative error in the last iteration. (c) Comparison of the convergence behavior of the three considered mathematical representation of Dirac delta function (Lorentzian, Gaussian, and unit pulse) versus temperature for three different mesh densities ($L_P$=23, $L_P$=31, and $L_P$=51) using SMRT approximation. (d) The impact of poor sampling of BZ on the thermal conductivity behavior versus temperature for the case of unit pulse representation of Dirac delta.

### 4.2. Impact of dispersion relations & BZ structure

In the previous subsection, we investigated the impact of the numerical implementation of our approach on the thermal conductivity prediction to single them out. For the rest of this section, only the results based on Lorentzian function representation of Dirac delta and a mesh with $L_p = 23$ are reported. Now, to discern the effect of dispersion models and relations, it is more constructive to start by examining the emerging phonon spectrum for each of them. Figure 6 shows phonon DOS for linear versus nonlinear dispersion relations of ISO1 (a) and FANISO (b) dispersion models. It is obvious that the linear dispersion assumption mitigates the effect of Van Hove singularity for the case of isotropic continuum, by reducing the height of the resolved peaks. For the case of FANISO model, linear dispersion increases the fraction of phonon modes belong to the low energy regime at the cost of the high energy modes. Similar trends are noted for isotropic continuum models but the effect is less pronounced.

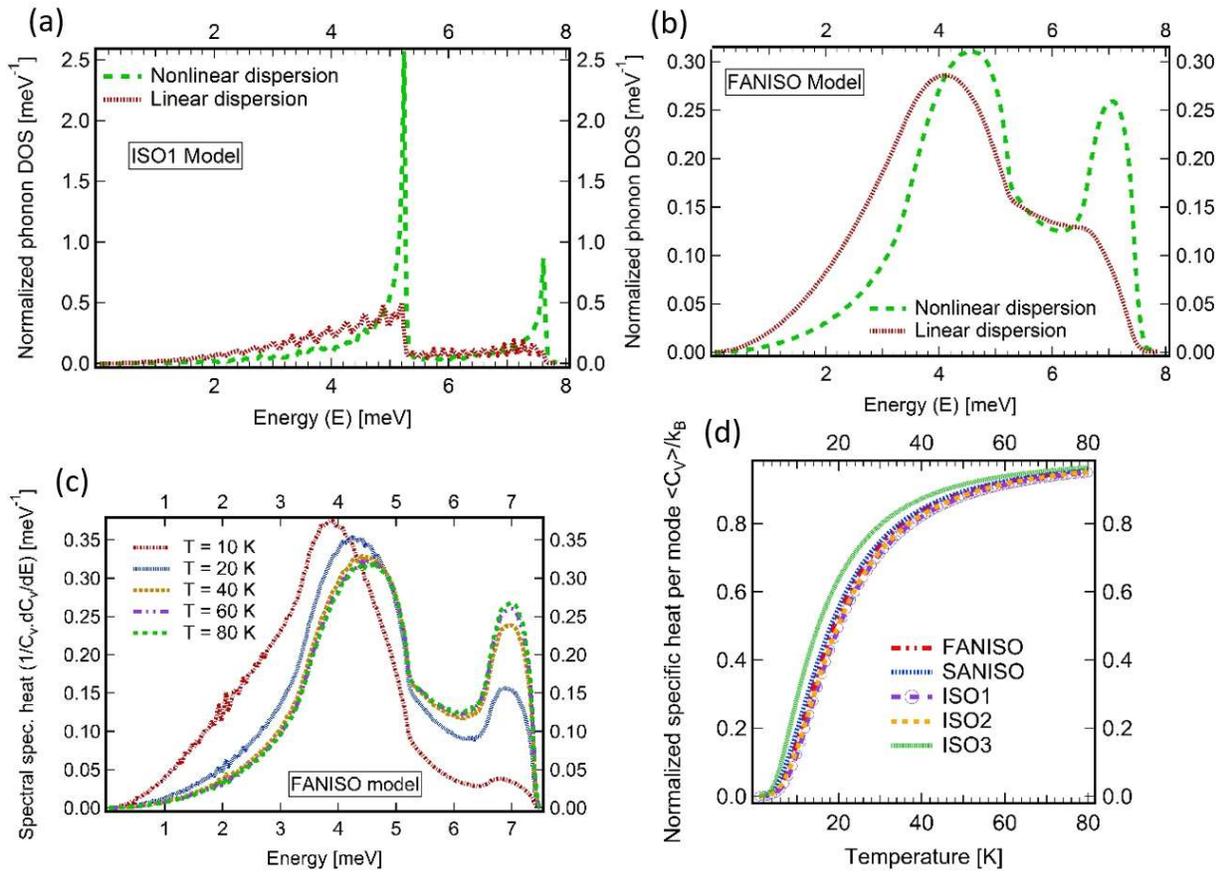

*Fig. 6. The impact of dispersion relation (linear versus nonlinear) on phonon Density Of States (DOS) for ISO1 (a) and FANISO (b) dispersion models. (c) Normalized spectral specific heat at different temperatures using FANISO dispersion model vs. energy. (d) The isochoric specific heat for different dispersion models normalized by the classical limit.*

Specific heat is one quantity that affect thermal conductivity, see Eq. (31). By studying the behavior of spectral specific heat profile, we can gain insight toward identifying which portion of phonon spectrum is the dominant heat carriers over temperature. The high collective contribution to specific heat of intermediate-energy phonon modes, at intermediate and high temperature range, can offset their relatively lower values of relaxation times with respect to low energy modes. Figures 6 (c) provides the spectral contribution to the harmonic isochoric specific heat versus energy at different temperatures using FANISO model. Apparently, the energy of the dominant contributing modes changes with temperature up to 20 K, as higher energy modes (with higher DOS) are getting occupied by phonons. For temperatures above 20 K, the two distinct maxima observed correspond to the ensemble of transverse and longitudinal modes near BZ surface, respectively, in accordance to phonon DOS. This is contrary to Debye model prediction, with one peak that is always located at normalized energy ($E/k_BT$) ~ 3.8, with all polarization branches lumped together. Figure 6(d) shows the impact of dispersion models on the harmonic isochoric specific heat (normalized by the classical limit), as a function of temperature, for FANISO model. These results depict that harmonic properties are less sensitive to the used dispersion model.

Figure 7 reveals relaxation times as a function of the wavevector in the three high symmetry directions for different polarization branches at 20 K for FANISO, SANISO and ISO1 dispersion models using SMRT approximation (a-c) and iterative scheme (d-f). In addition, relaxation times of normal and umklapp processes from FANISO model using SMRT approximation are shown (g-h). From this, we observe oscillations in relaxation time profile when isotropic continuum model is applied. These oscillations are damped, to some extent, when SANISO model was used. On the other hand, FANISO model produces fairly smooth profiles as a result of eliminating the roughness in phonon DOS due to Van Hove Singularity, which evidences the role played by dispersion data in low symmetry direction. On the other hand, relative values of the relaxation times of normal and umklapp processes indicates clearly that normal processes are the dominant scattering mechanism for most of phonon modes in the BZ, except for small fraction of longitudinal modes located towards BZ surface. So, applying SMRT approximation should be questionable. This observation holds true for other dispersion models as well. In Fig. 8, thermal conductivity profile as a function of temperature, from our three isotropic dispersion models (ISO1, ISO2, and ISO3) are plotted for both spherical BZ and truncated octahedron. In this regard, predictions based on Callaway's, SMRT, and umklapp relaxation times are compared. The order of magnitude difference between Callaway's and SMRT approximation prediction for thermal conductivity, over the entire temperature range, conforms well to our observation of the predominance of normal processes. The discrepancy between the different isotropic dispersion models, constructed from the same interatomic potential, leaves no doubt that isotropic continuum approximation is not adequate. In addition, comparing the prediction of umklapp based conductivity for the two BZ shapes shows how introducing a restricting approximation that phonon triplets will always be coplanar for spherical BZ, with pseudo-reciprocal lattice vector that is always selected to be collinear with the phonon wavevector that belongs to q`` space, produces unpredictable pattern. From Table 2, we know that BZ shape has negligible effect on $P_3$, however, anisotropy of umklapp processes in case of truncated octahedron with ISO3 dispersion mode, makes its pattern similar to the case of spherical BZ and ISO2 dispersion model with its non-degenerate transvers branches. Despite that, the overall impact of BZ shape on the thermal conductivity is minimal, due to the dominance of normal processes and hence dispersion model is the determining factor.

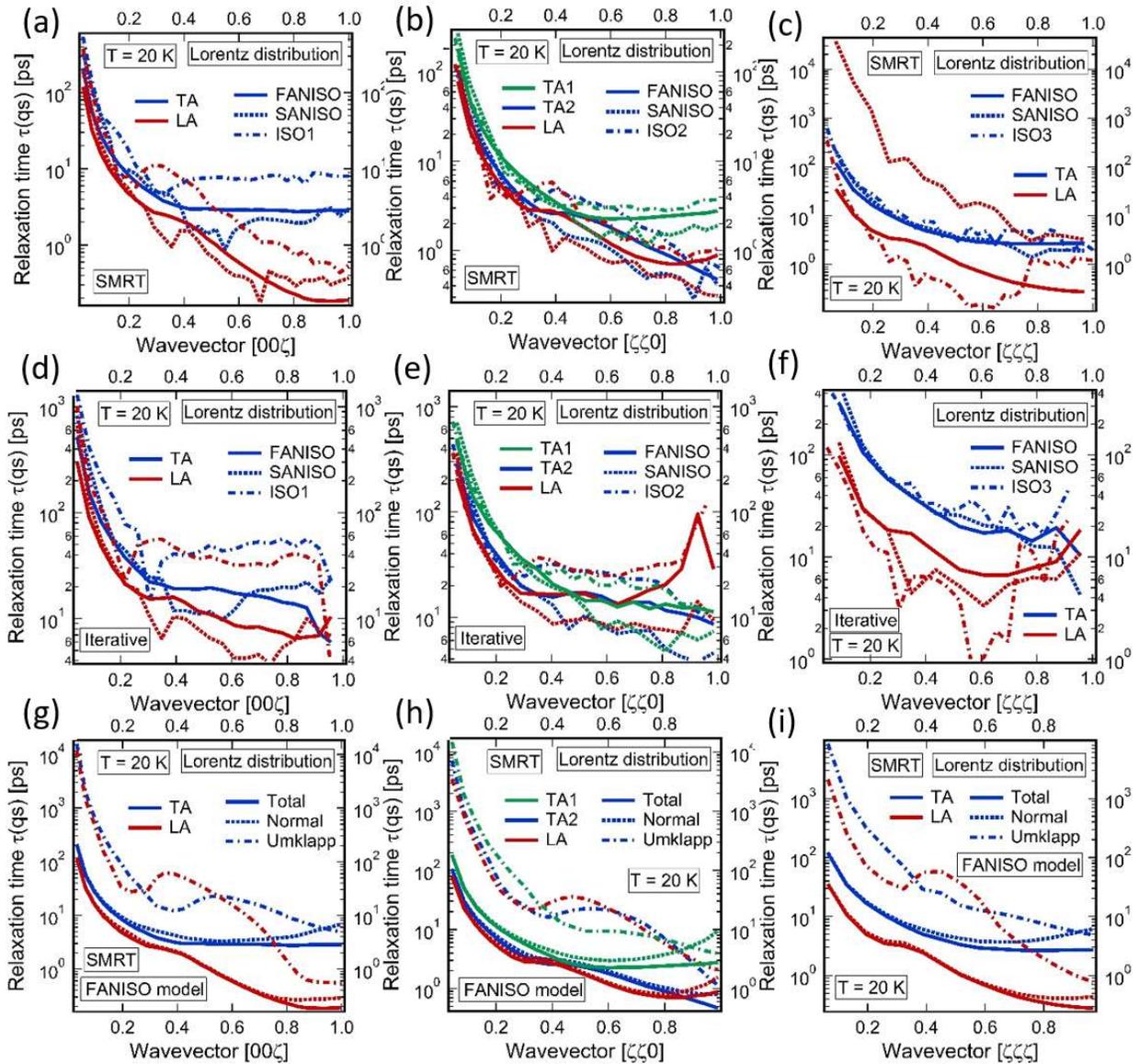

Fig. 7. Impact of dispersion model on Mode-dependent relaxation times in the three principal crystallographic directions ([001], [110], and [111]) at 20 K for different polarization branches using SMRT model (a-c) and iterative scheme (d-f). In addition, normal and umklapp processes relaxation times for different polarization branches in the same directions at 20 K, using FANISO dispersion model and SMRT approximation, are shown (g-i). Wavevectors are normalized by the maximum wavelength in each direction.

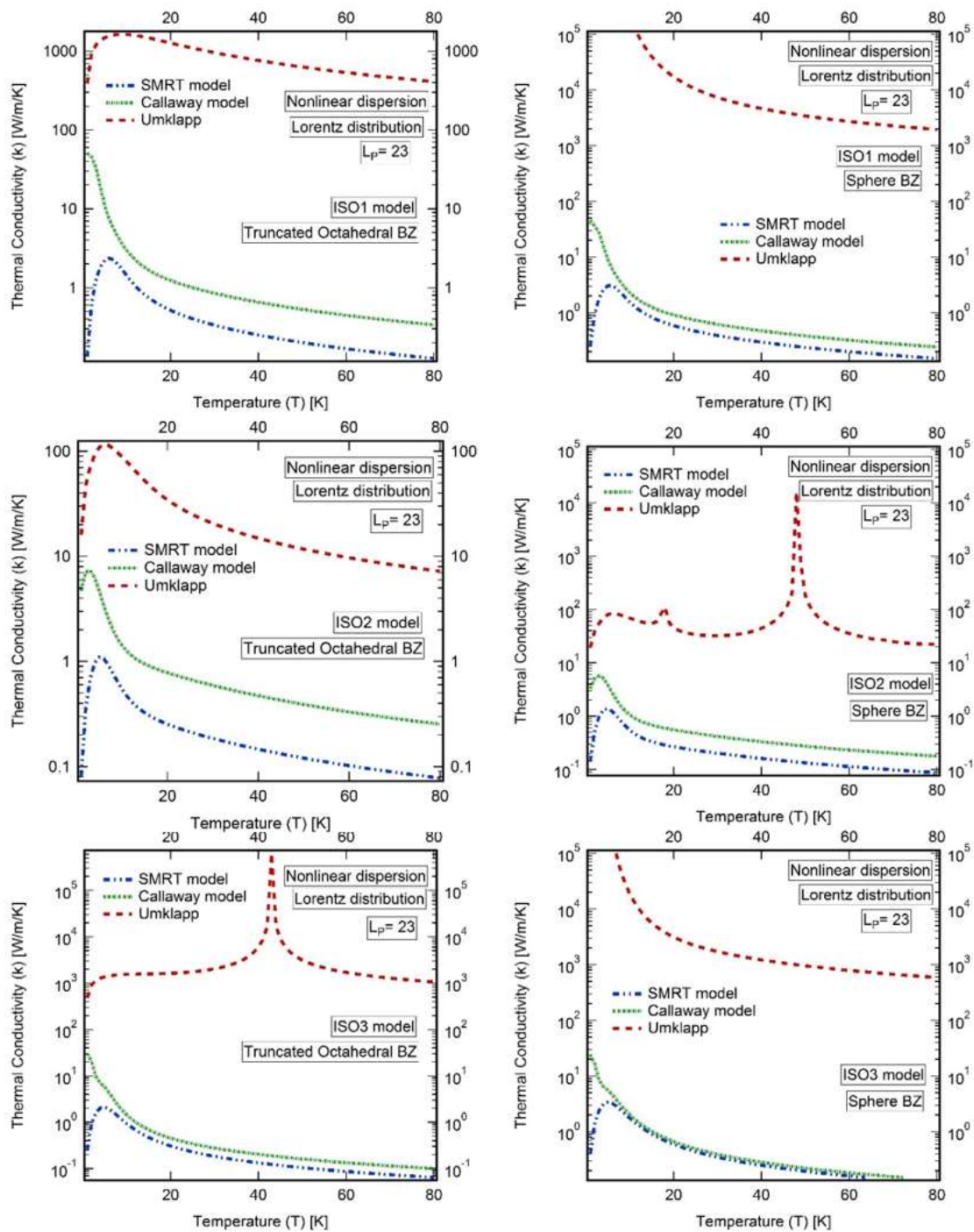

Fig. 8. BZ structure impact (truncated octahedron versus sphere) on thermal conductivity profile (as a function of temperature) for the three different isotropic dispersion models, namely, ISO1 ISO2, and ISO3, and for three different models of relaxation time: SMRT, Callaway, and Umklapp processes only (under SMRT approximation).

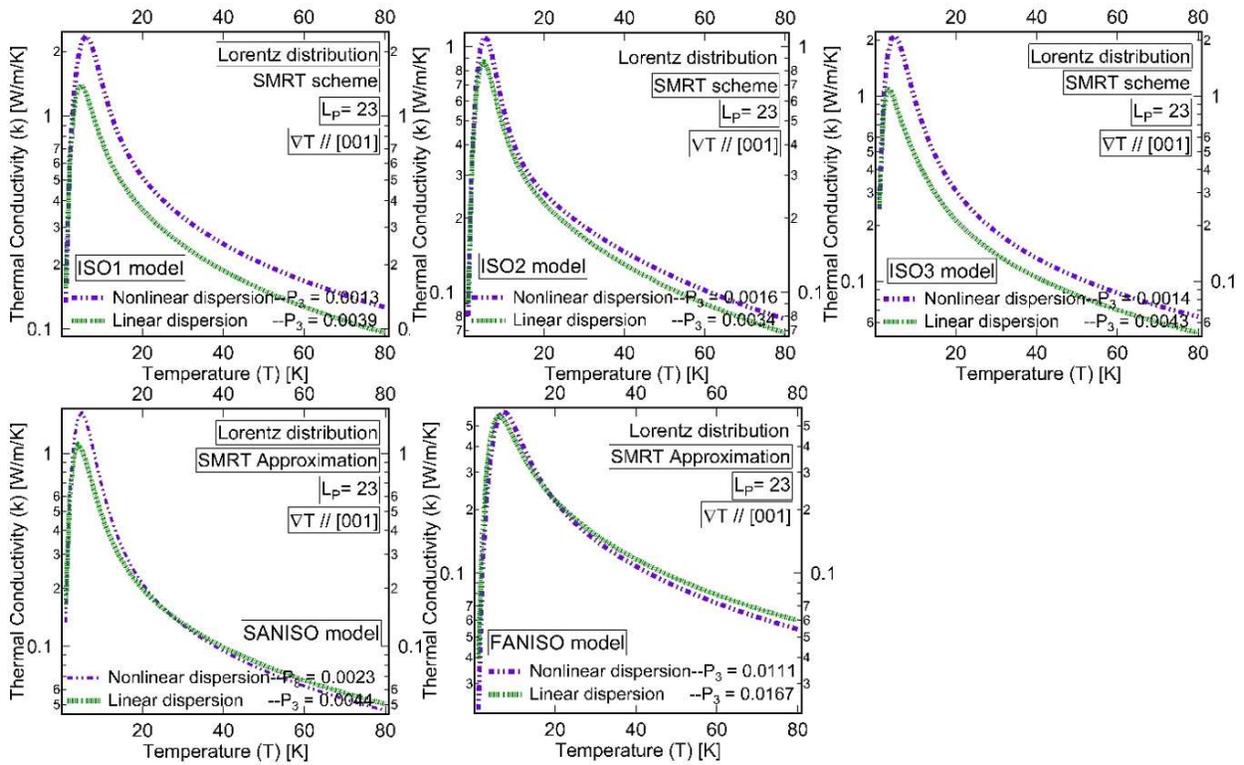

Fig. 9. The impact of dispersion relation (linear vs. nonlinear) on thermal conductivity profile (as a function of temperature) for ISO1, ISO2, ISO3, SANISO, and FANISO dispersion models using SMRT approximation.

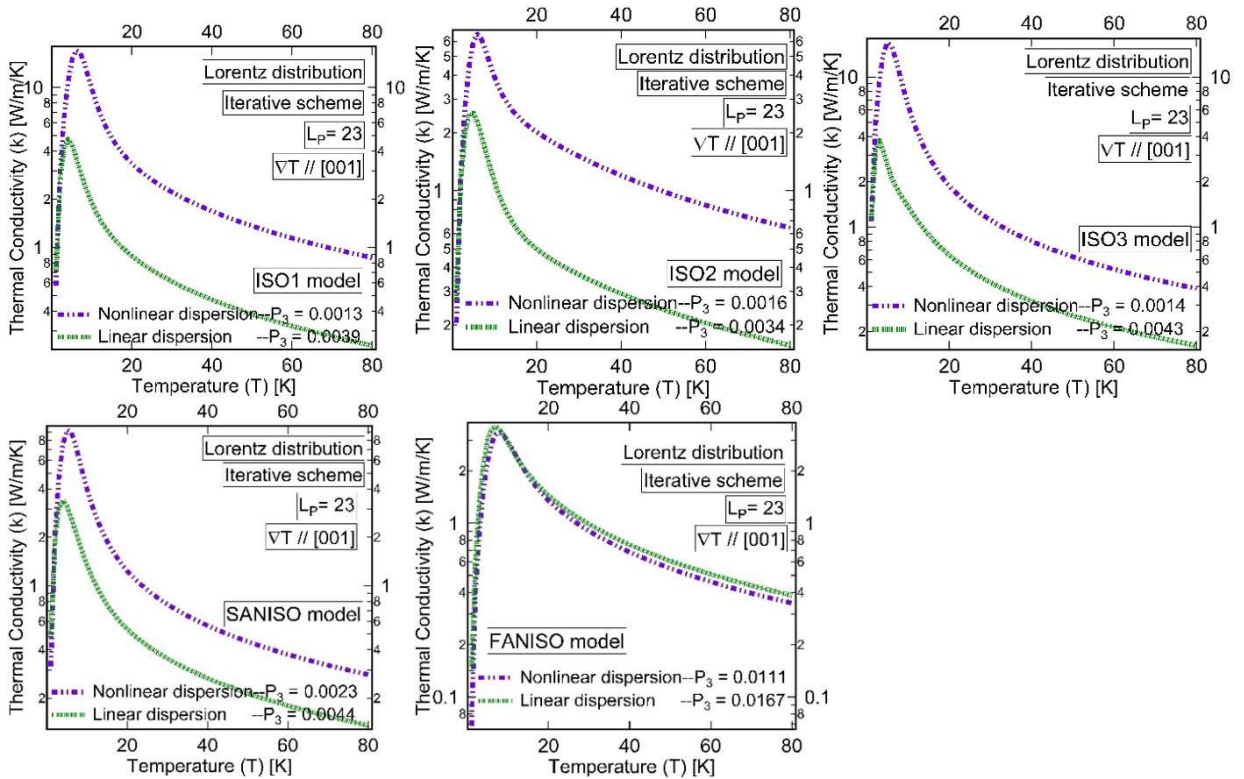

Fig. 10. The impact of dispersion relation (linear vs. nonlinear) on thermal conductivity profile (as a function of temperature) for different dispersion models using iterative scheme for relaxation time calculations.

Figures 9 and 10 compare between the thermal conductivity profiles of linear and nonlinear dispersion relations, for differet dispersion models, using SMRT approximation and iterative scheme, respectively. They evidences that the combined effect of accounting for dispersion anisotropy and assuming linear dispersion relation can counteract the effect of linear dispersion alone under isotropic continuum approximation. For example, while linear dispersion increases the available 3-phonon phase space, we can see clearly this does not always give lower values for thermal conductivity in case of SANISO and FANISO dispersion models. So, the overall behavior depends also on the dispersion models used, and some times on the temperature range under consideration.

### 4.3. Thermal conductivity anisotropy and relaxation times correlation effects

In Sec. 4.2 we quanitatively assesed the impact of the input parameters on the thermal conductivity, namely dispersion model and relations and the BZ structure. We also showed the dominance of normal scattering processes, which made SMRT approximation indaquate. In this subsection, we focus on the impact of relaxation time models and how taking in consideration the off-diagonal elements of phonon collision operator for umklapp processes reveals anisoropy in the thermal conductivity of cubic argon. Figures 11(a) and 11(b) compare thermal conductivity obtained from different dispersion models using nonlinear dispersion for the iterative scheme and SMRT approximation, respectively. Experimental data, shown in Fig. 11(a), compares very well with FANISO model prediction. Discrepancy at high temperature may be attributed to disregarding 4-phonon processes, which can play significant role in thermal resistivity of solid argon at high temperature, as predicted in Ref. [50]. However, without accounting for the effect of thermal expansion using, for example, temperature dependent dispersion curves, we are not able to confirm that. In addition, it is obvious that the impact of dispersion models on thermal conductivity, as a function of temperature, is not monotonic. For example, the same dispersion model can overestimate the conductivity at a certain temperature. In Sec. 2.3 it was indicated that, among the models considered, Srivastava's and the iterative relaxation times are the only ones that can predict anisotropy in thermal conductivity. It was also clarified that this anisotropic behavior within our approach, which considers linear regime, depends only on the direction of the applied temperature gadient, but not on its magnitude. Figure 12(a) presents thermal conductivity profile, as a function of temperature, for the different relaxation time models using FANISO dispersion model, when temperature gradient is applied along one of the three high symmetry directions, while Fig. 12(b) shows the results based on umklapp processes only, for both Srivastava's model and the iterative scheme. Since the differences in the values of thermal conductivity for different temperature gradient directions are small and exhibit temperature dependence, Fig. 12(c) provides the relative change of thermal conductivity, by taking the temperature gradient in [111] direction as the reference, for both Srivastava's model and the iterative scheme. Figure 12(d) is similar to Fig 12(a), but for SANISO dispersion model. These figures reveal that anisotropy in condutivity is persistent over the entire temperature range, however, this effect is negligible at temperatures higher than 10 K, where the absolute difference lies within the uncertainty of experimental measurements. Moreover, the anisotropy in thermal conductivity, due to phonon focusing, is more distinct for FANISO dispersion model. Strikingly, Srivastava's model prediction is completely different from the iterative scheme. While, for the iterative scheme, anisotropic conductivity is more

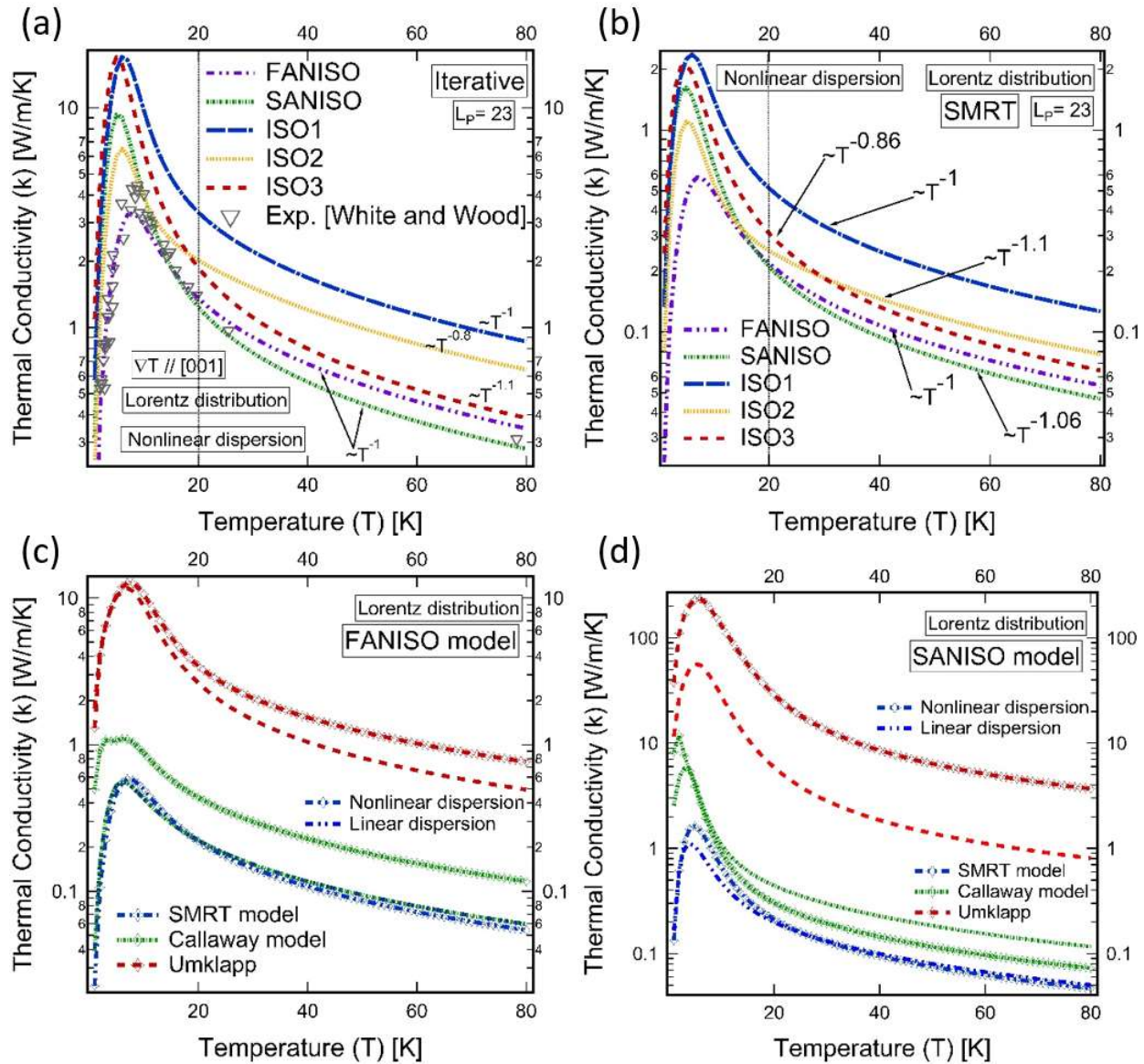

Fig. 11. (a) A comparison between thermal conductivity prediction for different dispersion model using iterative scheme and experimental data of argon (taken from White and Woods [51]). (b) Thermal conductivity prediction for different dispersion models under SMRT approximation. (c) and (d) Impact of dispersion relation on thermal conductivity prediction for three different models of relaxation time: SMRT, Callaway, and Umklapp processes only (under SMRT approximation) using FANISO and SANISO dispersion models, respectively.

Page | 37

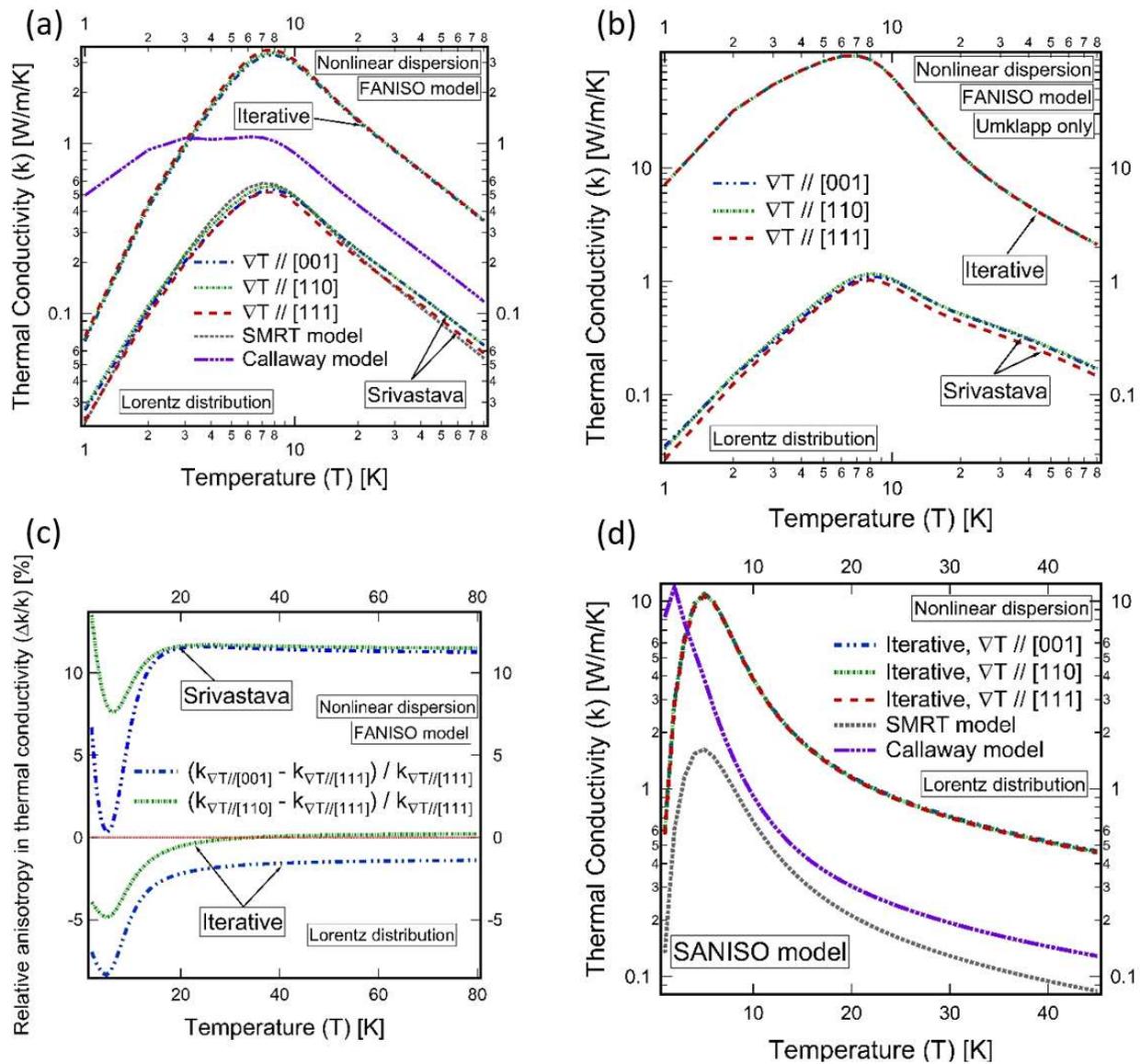

Fig. 12. Predicted anisotropy in thermal conductivity, when the temperature gradient is applied towards one of the three high symmetry directions [001], [110], and [111] respectively, for Srivastava model and the iterative scheme using FANISO dispersion model, along with Callaway's and SMRT models predictions (a). The impact of neglecting the normal processes contribution to thermal resistivity on the anisotropy of thermal conductivity (b). The relative percentage change in thermal conductivity, when the direction of applied temperature gradient changed from [111] direction, versus temperature for Srivastava model and the iterative scheme (c). Thermal conductivity anisotropy prediction for SANISO dispersion model using iterative scheme along with Callaway's and SMRT models predictions (d).

remarkable at low temperature with maximum relative changes below 8 % at the peak conductivity, which falls off to small values at high temperature, Srivastava's model produces anisotropic conductivity that levels off at relative change of 10 % at high temperatures. In addition, Srivastava's model predict the conductivity to be higher, when temperature gradient is applied along [110] direction, and gives minimal differences in high-temperature-conductivity when this direction is changed to [100]. On the contrary, the highest conducivity in iterative scheme is obtainend when the temperature gradient is parallel to [111] direction. Moreover, it shows that conductivity for [110] temperature gradient reaches a slightly higher values than [111] direction at high temperature. These discrepancies can be interpreted in light of the complete reliance of Srivastava's model on the anisotropy of the IBZ to account for the off-diagonal elements of the umklapp processes collision kernel (without making any explicit account to the dispersion data). This is obvious from Fig 12(b), with anisotropy more pronounced for Srivastava's model. The other factor is the statistical average tratment of the coupling constants in Srivastava's model, in contrast to the iterative scheme. This particular part is the main reason for the failure of Callaway's model to reproduce experimental data at the entire temperature range, in contrast to the iterative solution, however it captured the right order of magnitude, see Fig. 12(a).

To elaborate on the significance of how the relaxation time was statistically averaged out, for the treatment of correlation effects, mode-averaged relaxation time was calculated using two diferent weighting functions, namely DOS and the spectral specific heat. Figure 13(a) shows the mode-averaged relaxation time, as a function of temperature, for the total SMRT as well as the scattering rates for normal and umklap processes, using the two weighting functions. Effective mode-averaged relaxation times (weighted by DOS), as a function of temperature, for different relaxation time moels are displayed in Fig. 13(b), while Fig. 13(c) shows the mode-averaged Mean Free Path (MFP) for both iterative and SMRT models. As would be expected, the two different weighting functions yield the same mode-aevarged relaxation time at high temperature. Nevertheless, spectral specific heat-weighted relaxation times shows exponential increase at low temperature in similarity to DOS-weighted mode-averaged relaxation times due to Callaway's and Srivastava's models. On the other hand, DOS-weighted mode-averaged relaxation times due to SMRT and iterative models exhibit no temperature dependency at low temperature. From that, we suggest the use of DOS as the appropriate weighting function to evaluate mode-averaged relaxation time, contrary to Callawy's model. At the same time, the similarity between Callaway's and the iterative mode-averaged relaxation times at high temperature, in apparent contradiction to their discrepancies in thermal conductivity prediction, can be noticed. This evidences that treating coupling constants at individual mode level is critical for accurate representation of correlation effects. It is worth to mention that mode-averaged MFP follow the same pattern as the mode-averaged relaxation time.

Figure 14 shows the effective spectral relaxation time at two different temperatures (20 and 80 K) for normal and umklapp processes, as well as the total relaxation time using SMRT approximation and the iterative scheme, for FANISO dispersion model. The temperature gradient was taken parallel to [001] for the iterative solution. Again, it emphasizes the predominance of the normal processes over the whole spectrum, which makes SMRT approximation fail for that reason, for an appreciable fraction of the spectrum. SMRT relaxation times are smaller at 80 K than the minimum acceptable values for this model to be of physical significance. As a result, the predicted conductivity values are severely underestimated [52]. By referring to the exponents of the power function fit for different scattering processes, a temperature dependency was found. Although the values of energy exponents are within the suggested values in literature, showing temperature dependence raise concerns about the validity of applying Holland's model, which postulates the dependency of relaxation time on temperature and energy is separable.

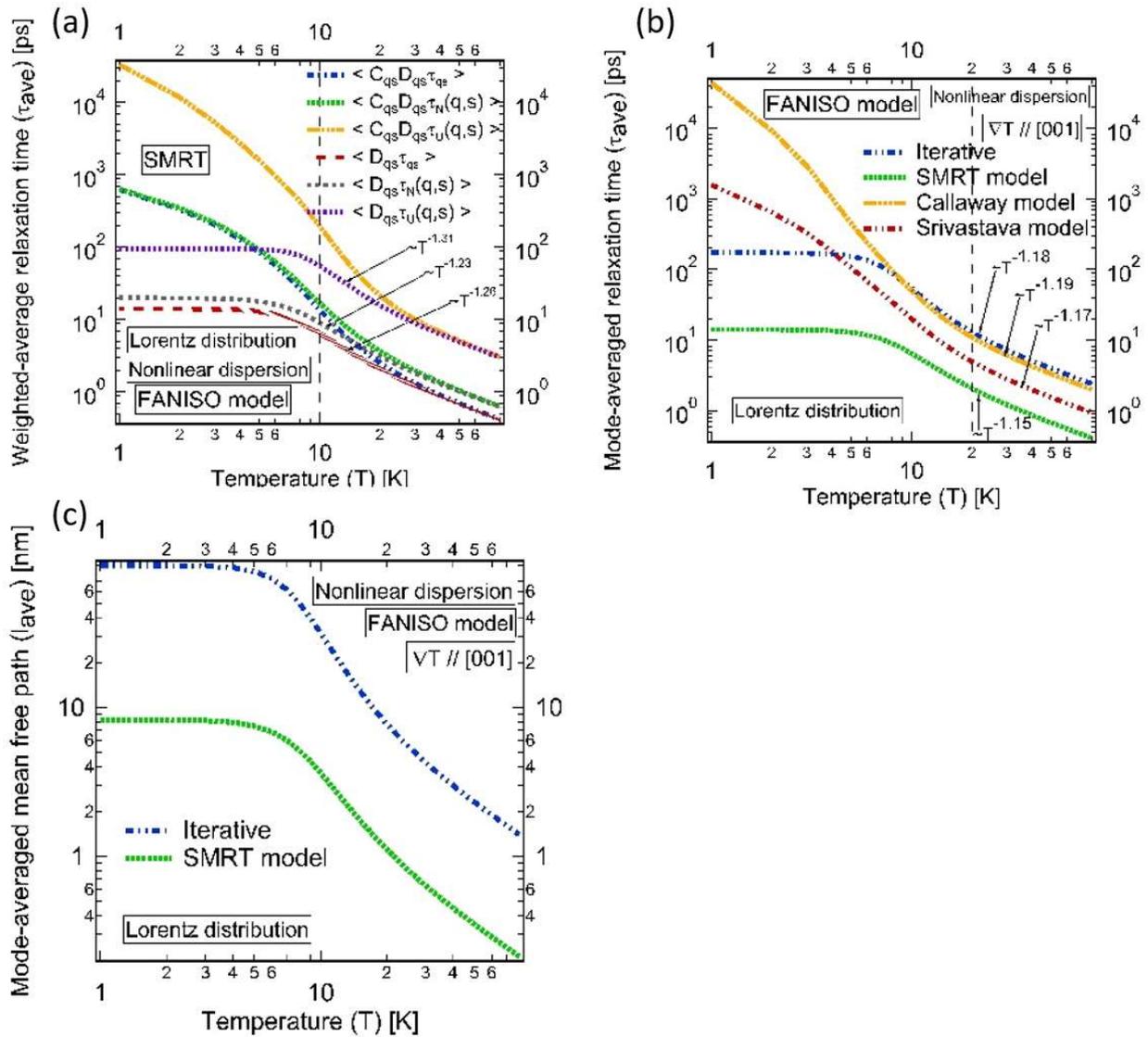

Fig. 13. Mode-averaged relaxation time as a function of temperature using two different weighting functions, namely, DOS and the product of DOS and specific heat, for normal, umklapp processes along with the total relaxation times under SMRT approximation (a). The DOS weighted Mode-averaged relaxation time for different relaxation time models (b). The DOS weighted Mode-averaged Mean Free Path (MFP) for SMRT model and iterative scheme(c).

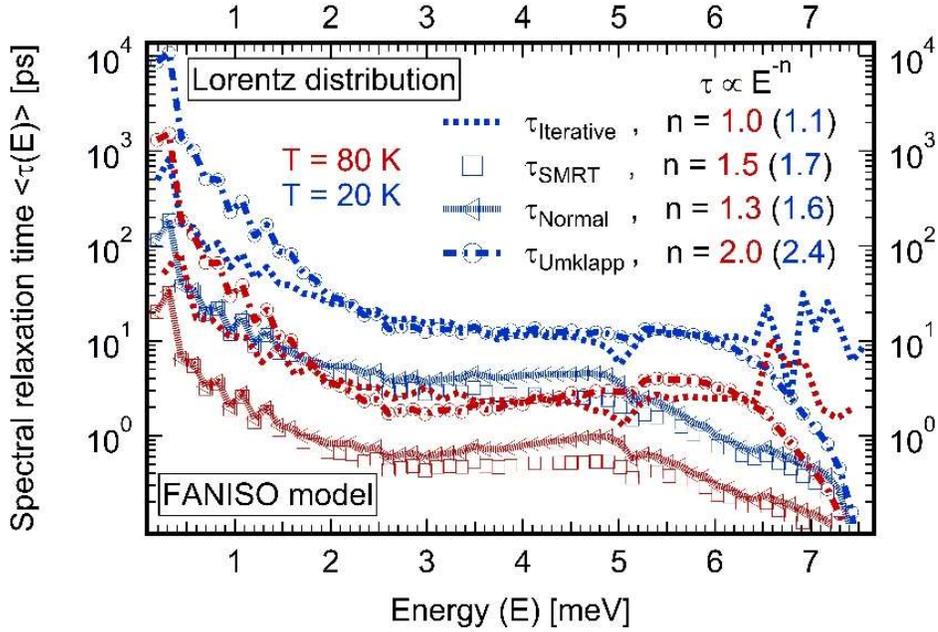

Fig. 14. Effective spectral relaxation time at two different temperatures (20 and 80 K) for normal and umklapp processes, as well as the total relaxation time using SMRT approximation and the iterative scheme, for FANISO dispersion model. The temperature gradient was taken parallel to [001] for the iterative solution.

Figure 15 shows color maps of the spectral thermal conductivity and the normalized spectral conductivity as a function of temperature for different polarization branches and the total one, using both SMRT approximation and the iterative scheme. The figure shows that, while SMRT approximation predicts low energy phonon modes to be the dominant heat carriers over the entire temperature range, accounting for the correlation between the scattering of different phonon modes by using the iterative solution shows that high energy modes contribution to thermal conductivity predominates at high temperature. This Figure supports MD study findings [17] of the existence of two distinct modes of heat propagations, due to incoherent scattering (dominating in low energy modes), and coherent scattering (dominating in high energy phonon modes). This evidences a crossover between the particle-like mode of phonon transport (phonon diffusion) to the wave-like mode at high temperature, as high energy phonon modes become the dominant heat carriers.

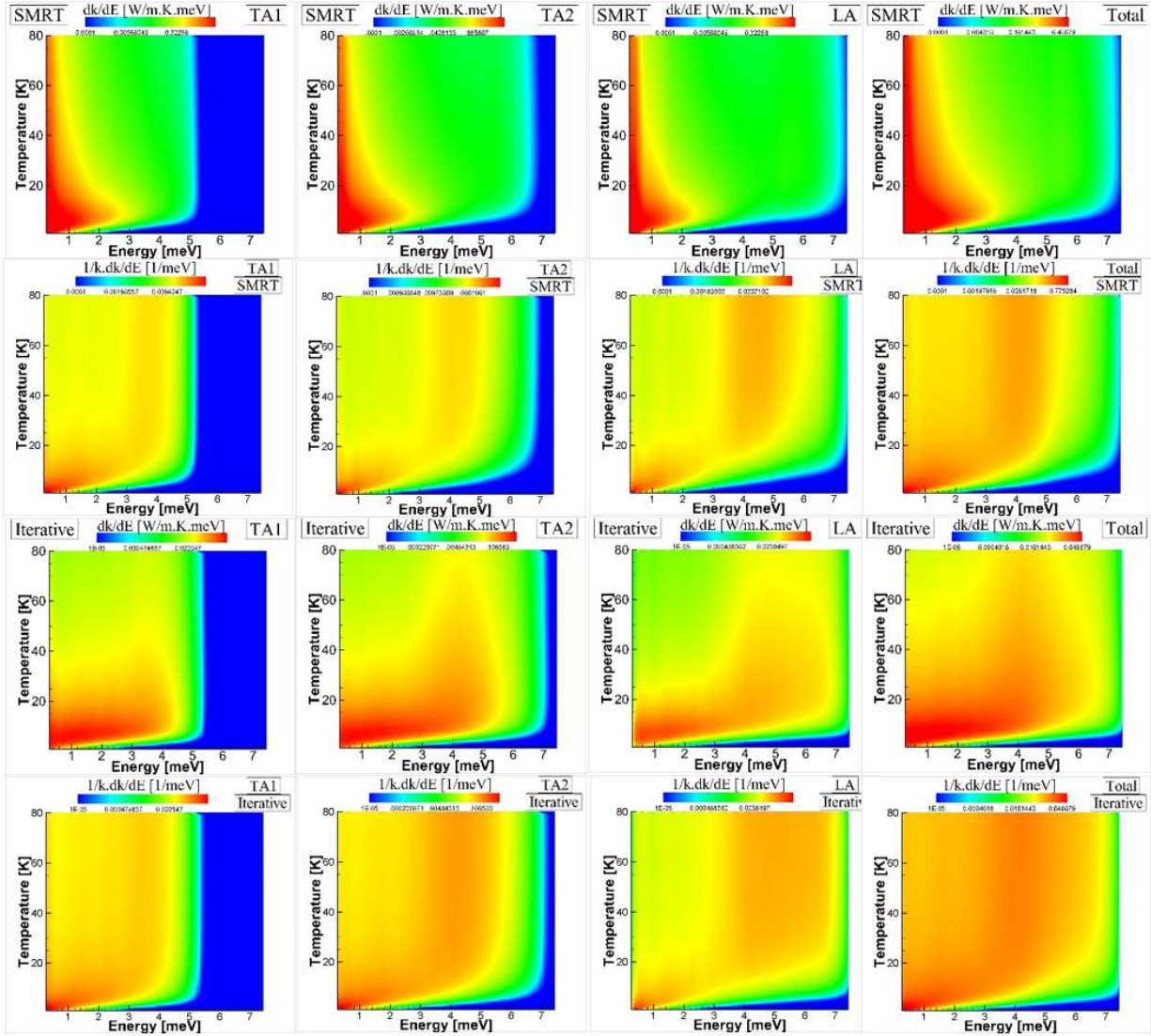

Fig. 15. Color map of spectral contribution to thermal conductivity, dk/dE, and normalized spectral conductivity, (1/k) dk/dE, versus temperature for the polarization branches (TA1, TA2, and LA) and their total contributions using both SMRT approximation and the iterative scheme. FANISO dispersion model, nonlinear dispersion, and Lorentz distribution were employed in these simulations. The temperature gradient was taken parallel to [001] for the iterative solution.

## 5. Summary and Outlook

A robust algorithm is presented for directly incorporating the dispersion curves of high symmetry directions in phonon transport calculations. In this algorithm, anisotropy of dispersion relations is accounted for by linearly interpolating dispersion data over the IBZ between high symmetry directions (FANISO dispersion model). The resulting model alleviates the need to solve the computationally expensive lattice dynamics problem for general wavevector in the BZ to find the associated phonon frequencies. We applied our model to Mie—Lennard-Jones argon. For accurate calculation of three-phonon scattering rates over the temperature range (2–80 K), our adjustable parameter-free representation of Dirac delta function was employed. A simple cubic mesh was used to tessellate the BZ (truncated octahedron) and special q-points were taken as sample points and the symmetry properties of FCC is exploited to reduce the computation domain to the IBZ, which yielded an efficient and rapidly converging scheme. This model was employed in conjunction with kinetic theory for thermal conductivity prediction. The contribution of the off-diagonal elements of phonon collision operator to the scattering strength was sought using iterative BTE solver. Spectral relaxation times, mode-averaged relaxation time, mean free paths, and spectral thermal conductivity over the whole temperature range were presented, as well as three-phonon scattering phase space.

The impact of several numerical approximations commonly encountered in solving the linearized BTE for phonons, within harmonic approximation—perturbation theory approach, are studied and benchmarked against our experimentally validated model. These include the dispersion relations, BZ structure, and effective relaxation time models. For dispersion relations, three isotropic continuum approximations, ISO1, ISO2, and ISO3, using dispersion curves of the [001], [110], and [111] directions, respectively, were considered. Another model for the anisotropy of nonlinear elastic medium (SANISO model) was also compared with FANISO model. The goal was to investigate quantitatively the impact of the dispersion model on calculated phonon frequency spectrum, and the sensitivity of the computed relaxation times and thermal conductivity to the used dispersion model. Linear dispersion (BZ center to edge) was compared with nonlinear dispersion (using trigonometric function). The effect of the anisotropy of umklapp processes was elucidated by using two different shapes of the BZ for FCC crystal. The exact shape of FCC BZ (truncated octahedron) with the well-defined reciprocal lattice vectors (independent of the interacting phonon triplets) was considered alongside with the spherical shape with pseudo-reciprocal lattice vector that is always selected to be collinear with the phonon wavevector that belongs to q`` space (thus adding a restricting approximation that phonon triplets will always be coplanar). Different available models for the treatment of the correlation between the relaxation of different phonon states to equilibrium in relaxation times calculations, namely, SMRT, Callaway's, and Srivastava's models, were also compared.

By investigating different dispersion models, 3-phonon phase space was found to be sensitive to the anisotropy of dispersion curves, as values lower by an order of magnitude were obtained for isotropic models. The underestimation of $P_3$ under isotropic continuum approximation is marginally compensated for by assuming linear dispersion relation. On the other hand, linear dispersion has a slight impact on $P_3$ in the case of FANISO model. Considering the overall impact of dispersion model and dispersion relation approximations on intrinsic lattice thermal conductivity, our results evidence that the overall behavior depends on the temperature range under consideration and in general cannot be qualitatively estimated, as error cancellations takes place. In addition, it was demonstrated that the departure from 1/T behavior of thermal conductivity at high temperature, under harmonic approximation, is partially because of isotropic continuum assumption. Accordingly, it is crucial to account for cubic anisotropy of dispersion curves for accurate thermal conductivity prediction, and the argument that employing isotropic continuum assumption, by using the dispersion curves in one crystallographic direction (most of the time [001] was chosen), is well-suited for cubic system has been quantitatively demonstrated to be flawed and unreliable. Although adopting the actual shape of BZ was revealed to be important for accurate umklapp scattering rates calculation, assuming spherical BZ did not remarkably influence the predicted thermal conductivity of FCC argon. This was elucidated by the dominance of normal processes over the entire temperature range. The importance of considering the correlation between the relaxation of different phonon modes at mode-level, and not in statistical average sense as Callaway's model does, was revealed. In addition, by illustrating the crossover between heat diffusion via particle-like phonons and wave-like propagation due to phonon collective scattering at high temperature, our results emphasized the essential role played by the collective relaxation of phonon modes. Furthermore, SMRT approximation was shown to significantly underestimate thermal conductivity for argon. Moreover, our approach predicted anisotropy in thermal conductivity of cubic argon. Finally, this study demonstrates the existence of characteristic peak thermal conductivity at finite temperature in perfect (defect free) crystals, regardless of the dispersion and relaxation times models used.

Several improvements could be supplemented to the computational model presented here in this study, which include considering the temperature dependence of the dispersion curves (quasi-harmonic approximation), mode-specific anharmonicity, employing higher order interpolation scheme for dispersion data in general directions, and incorporating the dispersion curves of additional crystallographic directions.

**Acknowledgements**

This material is partly based upon work supported as part of the Center for Materials Science of Nuclear Fuel, an Energy Frontier Research Center funded by the U.S. Department of Energy, Office of Sciences, Office of Basic Energy Sciences under award number FWP 1356, through subcontract number 00122223 at Purdue University, and partly by Idaho National Laboratory through a subcontract titled 'Microstructure Evolution in UO$_2$' at Purdue University.

**References**

[1] G.P. Srivastava, Tuning phonon properties in thermoelectric materials, Reports Prog. Phys. 78 (2015) 26501. doi:10.1088/0034-4885/78/2/026501.

[2] D.G. Cahill, P. V. Braun, G. Chen, D.R. Clarke, S. Fan, K.E. Goodson, P. Keblinski, W.P. King, G.D. Mahan, A. Majumdar, H.J. Maris, S.R. Phillpot, E. Pop, L. Shi, Nanoscale thermal transport. II. 2003-2012, Appl. Phys. Rev. 1 (2014) 11305. doi:10.1063/1.4832615.

[3] T. Feng, X. Ruan, Prediction of Spectral Phonon Mean Free Path and Thermal Conductivity with Applications to Thermoelectrics and Thermal Management: A Review, J. Nanomater. 2014 (2014) 1–25. doi:10.1155/2014/206370.

[4] T. Luo, G. Chen, Nanoscale heat transfer--from computation to experiment., Phys. Chem. Chem. Phys. 15 (2013) 3389–412. doi:10.1039/c2cp43771f.

[5] A.J. Minnich, Advances in the measurement and computation of thermal phonon transport properties, J. Phys. Condens. Matter. 27 (2015) 53202. doi:10.1088/0953-8984/27/5/053202.

[6] Z. Tian, S. Lee, G. Chen, Heat Transfer in Thermoelectric Materials and Devices, J. Heat Transfer. 135 (2013) 61605. doi:10.1115/1.4023585.

[7] M. Zebarjadi, K. Esfarjani, M.S. Dresselhaus, Z.F. Ren, G. Chen, Perspectives on thermoelectrics: from fundamentals to device applications, Energy Environ. Sci. 5 (2012) 5147. doi:10.1039/c1ee02497c.

[8] C.D. Kaddi, J.H. Phan, M.D. Wang, Computational nanomedicine: modeling of nanoparticle-mediated hyperthermal cancer therapy, Nanomedicine. 8 (2013) 1323–1333. doi:10.2217/nnm.13.117.

[9] S. Krishnan, P. Diagaradjane, S.H. Cho, Nanoparticle-mediated thermal therapy: Evolving strategies for prostate cancer therapy, Int. J. Hyperth. 26 (2010) 775–789. doi:10.3109/02656736.2010.485593.

[10] G.P. Srivastava, The Physics of Phonons, Taylor and Francis, New York, 1990.

[11] J.A. Reissland, The physics of phonons, Wiley, 1973.

[12] P.G. Klemens, Thermal conductivity and lattice vibrational modes, Solid State Phys. 7 (1958) 1–98.

[13] J.M. Ziman, Electrons and Phonons, Oxford University Press, London, 1960.

[14] O.N. Bedoya-Martínez, J.-L. Barrat, D. Rodney, Computation of the thermal conductivity using methods based on classical and quantum molecular dynamics, Phys. Rev. B. 89 (2014) 14303. doi:10.1103/PhysRevB.89.014303.

[15] J.E. Turney, E.S. Landry, A. J.H. McGaughey, C.H. Amon, Predicting phonon properties and thermal conductivity from anharmonic lattice dynamics calculations and molecular dynamics simulations, Phys. Rev. B - Condens. Matter Mater. Phys. 79 (2009) 1–12. doi:10.1103/PhysRevB.79.064301.

[16] A.J.H. McGaughey, M. Kaviany, Thermal conductivity decomposition and analysis using molecular dynamics simulations. Part I. Lennard-Jones argon, Int. J. Heat Mass Transf. 47 (2004) 1783–1798. doi:10.1016/j.ijheatmasstransfer.2003.11.002.

[17] H. Kaburaki, J. Li, S. Yip, Thermal conductivity of solid argon by classical molecular dynamics, Mater. Res. Soc. Symp. Proc. 538 (1999) 503.

[18] H. Kaburaki, J. Li, S. Yip, H. Kimizuka, Dynamical thermal conductivity of argon crystal, J. Appl. Phys. 102 (2007) 43514. doi:10.1063/1.2772547.

[19] L. Chaput, Direct solution to the linearized phonon boltzmann equation, Phys. Rev. Lett. 110 (2013) 1–5. doi:10.1103/PhysRevLett.110.265506.

[20] A. Chernatynskiy, S.R. Phillpot, Evaluation of computational techniques for solving the Boltzmann transport equation for lattice thermal conductivity calculations, Phys. Rev. B - Condens. Matter Mater. Phys. 82 (2010) 1–17. doi:10.1103/PhysRevB.82.134301.

[21] A.J.H. McGaughey, M. Kaviany, Quantitative validation of the Boltzmann transport equation phonon thermal conductivity model under the single-mode relaxation time approximation, Phys. Rev. B. 69 (2004) 1–12. doi:10.1103/PhysRevB.69.094303.

[22] P.A.M. Dirac, On the Theory of Quantum Mechanics, Proc. R. Soc. A Math. Phys. Eng. Sci. 112 (1926) 661–677. doi:10.1098/rspa.1926.0133.

[23] D.C. Wallace, Thermodynamics of Crystals, Dover Publications, INC, New York, 1998.

[24] Y.-J. Han, P.G. Klemens, Anharmonic thermal resistivity of dielectric crystals at low temperatures, Phys. Rev. B. 48 (1993) 6033–6042.

[25] J. Callaway, Model for lattice Thermal Conductivity at Low Temperatures, Phys. Rev. 113 (1959) 1046–1051. doi:10.1103/PhysRev.113.1046.

[26] M. Omini, A. Sparavigna, Beyond the isotropic-model approximation in the theory of thermal conductivity, Phys. Rev. B. 53 (1996) 9064–9073. doi:10.1103/PhysRevB.53.9064.

[27] A. Cepellotti, N. Marzari, Thermal Transport in Crystals as a Kinetic Theory of Relaxons, Phys. Rev. X. 6 (2016) 41013. doi:10.1103/PhysRevX.6.041013.

[28] M. Holland, Analysis of Lattice Thermal Conductivity, Phys. Rev. 132 (1963) 2461–2471. doi:10.1103/PhysRev.132.2461.

[29] J.D. Chung, A.J.H. McGaughey, M. Kaviany, Role of Phonon Dispersion in Lattice Thermal Conductivity Modeling, J. Heat Transfer. 126 (2004) 376. doi:10.1115/1.1723469.

[30] G.K. Horton, J.W. Leech, On the Statistical Mechanics of the Ideal Inert Gas Solids, Proc. Phys. Soc. 82 (1963) 816–854. doi:10.1088/0370-1328/82/6/302.

[31] B. Yang, G. Chen, Partially coherent phonon heat conduction in superlattices, Phys. Rev. B. 67 (2003) 1–4. doi:10.1103/PhysRevB.67.195311.

[32] J. Garg, G. Chen, Minimum thermal conductivity in superlattices: A first-principles formalism, Phys. Rev. B - Condens. Matter Mater. Phys. 87 (2013) 1–5. doi:10.1103/PhysRevB.87.140302.

[33] Y. Wang, H. Huang, X. Ruan, Decomposition of coherent and incoherent phonon conduction in superlattices and random multilayers, Phys. Rev. B - Condens. Matter Mater. Phys. 90 (2014) 48–50. doi:10.1103/PhysRevB.90.165406.

[34] N. Zuckerman, J.R. Lukes, Atomistic visualization of ballistic phonon transport, in: Proc. HT2007, ASME, Vancouver, British Columbia, CANADA, 2007: pp. 1–9.

[35] A.G. Every, W. Sachse, K.Y. Kim, M.O. Thompson, Phonon focusing and mode-conversion effects in silicon at ultrasonic frequencies, Phys. Rev. Lett. 65 (1990) 1446–1449. doi:10.1103/PhysRevLett.65.1446.

[36] N.R. Werthamer, Self-consistent phonon formulation of anharmonic lattice dynamics, Phys. Rev. B. 1 (1970) 572–581. doi:10.1103/PhysRevB.1.572.

[37] H.R. Glyde, M.G. Smoes, Phonons in solid argon, Phys. Rev. B. 22 (1980) 6391–6402. doi:10.1103/PhysRevB.22.6391.

[38] A.K. Ghatak, L.S. Kothari, An introduction to lattice dynamics, Addison-Wesley, 1972.

[39] P.B. Allen, Improved Callaway model for lattice thermal conductivity, Phys. Rev. B - Condens. Matter Mater. Phys. 88 (2013) 1–5. doi:10.1103/PhysRevB.88.144302.

[40] L. Wei, P.K. Kuo, R.L. Thomas, T.R. Anthony, W.F. Banholzer, A. Inyushkin, A. Taldenkov, V. Ozhogin, K. Itoh, E. Haller, Thermal conductivity of isotopically modified single crystal diamond, Phys. Rev. Lett. 70 (1993) 3764–3767. doi:10.1103/PhysRevLett.70.3764.

[41] P. Torres, A. Torelló, J. Bafaluy, J. Camacho, X. Cartoixà, F.X. Alvarez, First principles kinetic-collective thermal conductivity of semiconductors, Phys. Rev. B - Condens. Matter Mater. Phys. 95 (2017) 1–7. doi:10.1103/PhysRevB.95.165407.

[42] L. Lindsay, D. a Broido, Three-phonon phase space and lattice thermal conductivity in semiconductors, J. Phys. Condens. Matter. 20 (2008) 165209. doi:10.1088/0953-8984/20/16/165209.

[43] J.W.L. Pang, W.J.L. Buyers, A. Chernatynskiy, M.D. Lumsden, B.C. Larson, S.R. Phillpot, Phonon lifetime investigation of anharmonicity and thermal conductivity of $UO_2$ by neutron scattering and theory, Phys. Rev. Lett. 110 (2013) 1–5. doi:10.1103/PhysRevLett.110.157401.

[44] A. Hamed, A. El-azab, Peak intrinsic thermal conductivity in non-metallic solids and new interpretation of experimental data for argon, J. Phys. Commun. 2 (2018) 15022. http://iopscience.iop.org/article/10.1088/2399-6528/aaa36f/pdf.

[45] W.R. Deskins, A. Hamed, T. Kumagai, C.A. Dennett, J. Peng, M. Khafizov, D. Hurley, A. El-Azab, Thermal conductivity of $ThO_2$: Effect of point defect disorder, J. Appl. Phys. 129 (2021). doi:10.1063/5.0038117.

[46] A. Hamed, Computational modeling of thermal transport using spectral phononBoltzmann transport equation, Ph.D. dissertation (Purdue University, 2017).

[47] G.K. Horton, H. Schiff, On the evaluation of equivalent Debye temperatures and related problems, Proc. R. Soc. A Math. Phys. Eng. Sci. 250 (1959) 248–265. doi:10.1098/rspa.1983.0054.

[48] W. V. Houston, Normal vibrations of a crystal lattice, Rev. Mod. Phys. 20 (1948) 161–165. doi:10.1103/RevModPhys.20.161.

[49] D.T. Morelli, G.A. Slack, High Lattice Thermal Conductivity Solids, in: High Therm. Conduct. Mater., Springer-Verlag, New York, 2006: pp. 37–68. doi:10.1007/0-387-25100-6_2.

[50] T. Feng, X. Ruan, Quantum mechanical prediction of four-phonon scattering rates and reduced thermal conductivity of solids, Phys. Rev. B. 93 (2016) 45202. doi:10.1103/PhysRevB.93.045202.

[51] G.K. White, S.B. Woods, Thermal Conductivity of Solid Argon at Low Temperatures, Nature. 177 (1956) 851–852.

[52] V.A. Konstantinov, Manifestation of the Lower Limit to Thermal Conductivity in the Solidified Inert Gases, J. Low Temp. Phys. 122 (2001) 459–465. doi:10.1023/A:1004877607357.

[53] H.J. Monkhorst, J.D. Pack, special points for Brillouin-zone integrations, Phys. Rev. B. 13 (1977) 5188–5192. doi:10.1103/PhysRevB.16.1748.

[54] A.H. MacDonald, S.H. Vosko, P.T. Coleridge, Extensions of the tetrahedron method for evaluating spectral properties of solids, J. Phys. C Solid State Phys. 12 (1979) 2991–3002. doi:10.1088/0022-3719/12/15/008.

[55] G. Karlowatz, Advanced Monte Carlo Simulation for Semiconductor Devices, Vienna University of Technology, 2009. http://www.iue.tuwien.ac.at/phd/karlowatz/diss.html (accessed June 15, 2017).

# Appendix A

This appendix provides the details of scheme used in this study for BZ discretization and BZ sums. This scheme was implemented into an in-house C++ code, with both serial and parallel versions implemented. Spectral decomposition technique was used for parallelization. Figure A.1. shows the FCC BZ (truncated octahedron) as well as its Irreducible wedge (IBZ). The coordinates of the vertices and the planes surrounding the IBZ are also given. Exploiting symmetry of cubic crystal structure, the computational domain was limited to the IBZ. Cubical mesh, which conforms with the symmetry of the cubic crystal structure, where employed. In addition, volume weighted averaging was applied, to calculate the number of phonon modes enclosed in each sub-grid. Different approaches have been used to evaluate the summation (integration) of spectral properties over the BZ, for example [23,53,54]:

- Discretization of IBZ (e.g. by octree method) and then interpolation between the known points using tetrahedron method (e.g. delaunay scheme).
- Random q points (which is less common now).
- Special q-points, including:
    - Chadi-Cohen scheme.
    - (Extended) Cunningham q-points.
    - Equi-distance or Monkhorst-Pack grid.

The procedures for sample points generation and summing over the BZ is explained below.

Phonons normal modes are homogenously distributed over the BZ, and for any function of phonon frequency we can replace the complete zone sum with partial zone sum by using the weighting factor $X_q$, which represents the volume fraction of the kth cube enclosed in IBZ. Eq. (A1) is used in all computations in this study that involved BZ sums for any mode-specific quantities. The symbol $\sum_{q}^{*}$ in this equation indicates that the summation is limited to the IBZ.

$$(3sN_o)^{-1}\sum_{qs} f(\omega(qs)) \rightarrow \frac{\sum_{s}\sum_{q}^{*} X_q f(\omega(qs))}{3s \sum_{q}^{*} X_q} \quad (A1)$$

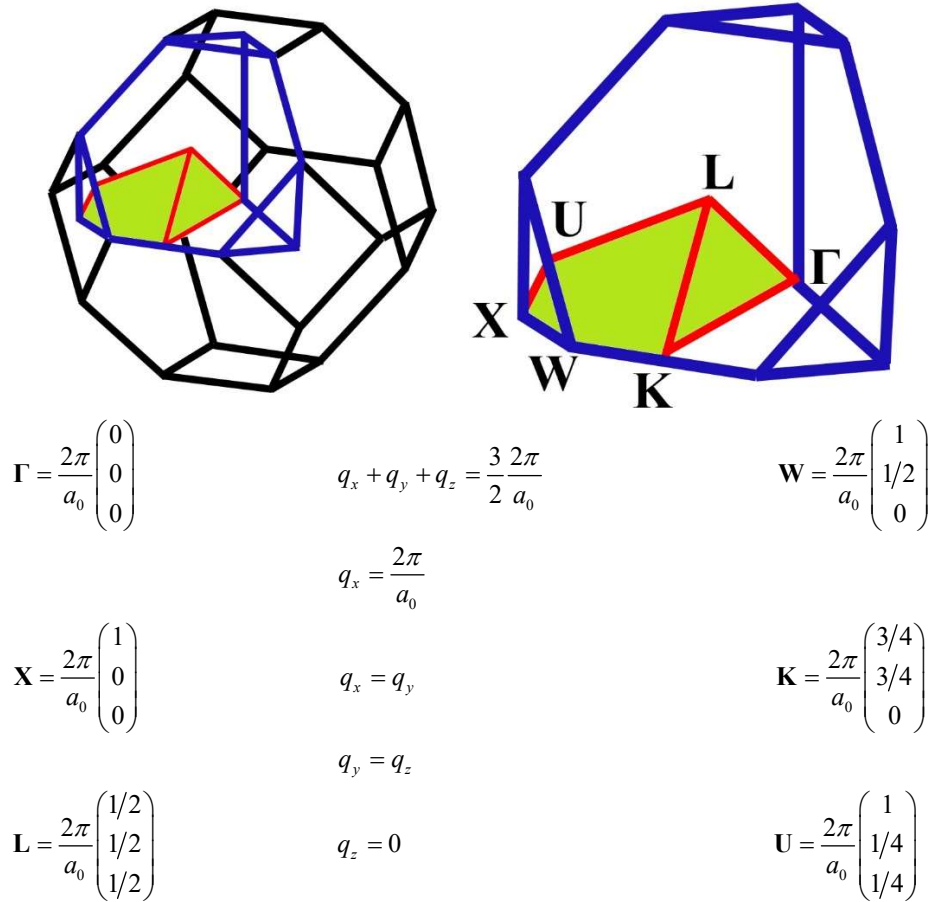

$$\Gamma = \frac{2\pi}{a_0}\begin{pmatrix}0\\0\\0\end{pmatrix} \qquad q_x + q_y + q_z = \frac{3}{2}\frac{2\pi}{a_0} \qquad W = \frac{2\pi}{a_0}\begin{pmatrix}1\\1/2\\0\end{pmatrix}$$

$$q_x = \frac{2\pi}{a_0}$$

$$X = \frac{2\pi}{a_0}\begin{pmatrix}1\\0\\0\end{pmatrix} \qquad q_x = q_y \qquad K = \frac{2\pi}{a_0}\begin{pmatrix}3/4\\3/4\\0\end{pmatrix}$$

$$q_y = q_z$$

$$L = \frac{2\pi}{a_0}\begin{pmatrix}1/2\\1/2\\1/2\end{pmatrix} \qquad q_z = 0 \qquad U = \frac{2\pi}{a_0}\begin{pmatrix}1\\1/4\\1/4\end{pmatrix}$$

Figure A.1. FCC Brillouin Zone (BZ) and Irreducible BZ (IBZ) alongside with the planes forming IBZ and its vertices [23,55].

This scheme takes a representative set of q wavevectors distributed uniformly throughout the zone. A simple cubic lattice is convenient for FCC, and defined by three integers $p_1$, $p_2$, $p_3$ according to: $q(p) = (2\pi/L_P a)(P_1 \hat{x} + p_2 \hat{y} + p_3 \hat{z})$. Where, a is the lattice constant, and $L_p$ is a positive integer which defines the limits to be placed on $p_1$, $p_2$, and $p_3$. The mesh density is determined by this integer alone. For IBZ, we can generate these integers according to the following limits:

- $0 \leq p_3 \leq L_p$;
- $0 \leq p_2 \leq L_2$, where $L_2$ = minimum of ($p_3$, $1.5L_p - p_3$);
- $0 \leq p_1 \leq L_1$, where $L_1$ = minimum of ($p_2$, $1.5L_p - p_3 - p_2$).

For FCC, the real space lattice vectors R(N) are given by: $R(N) = (a/2)(N_1 \hat{x} + N_2 \hat{y} + N_3 \hat{z})$. Where; $N_1$, $N_2$, and $N_3$ are positive integers. We can easily show that: $q(p) \cdot R(N) = (\pi/L_p)(p_1 N_1 + P_2 N_2 + P_3 N_3)$.

For IBZ portion of FCC, the weighting factors can be assigned as follows:

- Within the volume:    $X_q = 1$;
- On any face:          $X_q = 1/2$;
- On the edge ΓL:       $X_q = 1/6$;
- On the edge ΓX:       $X_q = 1/8$;
- On all other edges:   $X_q = 1/4$;
- At Γ point:           $X_q = 1/48$;
- At X point:           $X_q = 1/16$;
- At L point:           $X_q = 1/12$;
- At U, W, and K points: $X_q = 1/8$.

The following procedure can be used to generate these weighting factors by carrying out the following tests on $p_1$, $p_2$, and $p_3$, which define the q vector:

- Test 1. If $p_1 = p_2 = p_3 = 0$, $X_q = 1/48$.
- Test 2. If test 1 is not satisfied:
  - If $p_1 = p_2 = 0$, $p_3 \neq L_p$, $X_q = 1/8$;
  - If $p_1 = p_2 = 0$, $p_3 = L_p$, $X_q = 1/16$.
- Test 3. If neither test 1 nor test 2 satisfied:
  - If $p_1 = p_2 = p_3 \neq \frac{1}{2} L_p$, $X_q = 1/6$;
  - If $p_1 = p_2 = p_3 = \frac{1}{2} L_p$, $X_q = 1/12$.
- Test 4. If none of tests 1, 2, or 3 is satisfied, set $X_q = 1$ and then multiply by a factor of half (1/2) every time one of the following five tests is satisfied:
  - $p_1 + p_2 + p_3 = 1.5 L_p$;
  - $p_3 = L_p$;
  - $p_1 = p_2$;
  - $p_1 = 0$;
  - $p_3 = p_2$.

Finally, we define the weighting factor using

$$w_i = \frac{X_i}{\sum_{q=1}^{M} X_q}.  \qquad (A2)$$